\input{psfig}
\documentstyle[12pt,epsf,eqsection]{article}

\begin{document}

\font\german=eufm10 scaled\magstep1     % 12-point Euler Fraktur (German)
\def\germang{\hbox{\german g}}
\def\germansu{\hbox{\german su}}
\def\germanso{\hbox{\german so}}

\font\sevenrm  = cmr7 
\def\fancyplus{\hbox{+\kern-6.65pt\lower3.2pt\hbox{\sevenrm --}%
\kern-4pt\raise4.6pt\hbox{\sevenrm --}%
\kern-9pt\raise0.65pt\hbox{$\tiny\vdash$}%
\kern-2pt\raise0.65pt\hbox{$\tiny\dashv$}}}
\def\fancycross{\hbox{$\times$\kern-9.8pt\raise3.6pt\hbox{$\tiny \times$}%
\kern-1.2pt\raise3.6pt\hbox{$\tiny \times$}%
\kern-10.5pt\lower1.0pt\hbox{$\tiny \times$}%
\kern-1.2pt\lower1.0pt\hbox{$\tiny \times$}}}
\def\fancysquare{%
\hbox{$\small\Box$\kern-9.6pt\raise7.1pt\hbox{$\vpt\backslash$}%
\kern+3.9pt\raise7.1pt\hbox{$\vpt /$}%
\kern-11.5pt\lower3.0pt\hbox{$\vpt /$}%
\kern+3.9pt\lower3.0pt\hbox{$\vpt\backslash$}}}
\def\fancydiamond{\hbox{$\diamond$\kern-4.25pt\lower3.8pt\hbox{$\vpt\vert$}%
\kern-2.25pt\raise7pt\hbox{$\vpt\vert$}%
\kern-7.05pt\raise1.57pt\hbox{\vpt --}%
\kern+5.25pt\raise1.57pt\hbox{\vpt --}}}

\newcommand{\taum}{\tau_{int,\vec{\cal M}}}
\newcommand{\taux}{\tau_{int,{\cal M}^2}}
\newcommand{\tauA}{\tau_{int,A}}
\newcommand{\taue}{\tau_{int,{\cal E}}}
\newcommand{\taudele}{\tau_{int,({\cal E}-\overline{E})^2}}

\newcommand{\tauxexp}{\tau_{exp,{\cal M}^2}}
\newcommand{\tauxinit}{\tau_{init,{\cal M}^2}}

\newcommand{\plotdot}{\makebox(0,0){$\bullet$}}
\newcommand{\plotcross}{\makebox(0,0){{\Large $\times$}}}

\newcommand{\plota}{\makebox(0,0){$\circ$}}      % 512^2 lattice
\newcommand{\plotb}{\makebox(0,0){$\star$}}      % 256^2 lattice
\newcommand{\plotc}{\makebox(0,0){$\bullet$}}    % 128^2 lattice
\newcommand{\plotd}{\makebox(0,0){{\scriptsize $+$}}}       % 64^2  lattice
\newcommand{\plote}{\makebox(0,0){{\scriptsize $\times$}}}  % 32^2  lattice
\newcommand{\plotf}{\makebox(0,0){$\ast$}}       % 16^2  lattice

\newcommand{\plotA}{\makebox(0,0){$\triangleleft$}}   % 512^2 lattice
\newcommand{\plotB}{\makebox(0,0){$\triangleright$}}  % 256^2 lattice
\newcommand{\plotC}{\makebox(0,0){$\diamond$}}        % 128^2 lattice
\newcommand{\plotD}{\makebox(0,0){{\scriptsize $\oplus$}}} % 64^2  lattice
\newcommand{\plotE}{\makebox(0,0){{\scriptsize $\otimes$}}}% 32^2  lattice
\newcommand{\plotF}{\makebox(0,0){{\scriptsize $\ominus$}}}% 16^2  lattice

\def\reff#1{(\ref{#1})}
\newcommand{\be}{\begin{equation}}
\newcommand{\ee}{\end{equation}}
\newcommand{\<}{\langle}
\renewcommand{\>}{\rangle}
\newcommand{\half}{ {{1 \over 2 }}}
\newcommand{\quarter}{ {{1 \over 4 }}}
\newcommand{\fourth}{\quarter}
\newcommand{\eighth}{ {{1 \over 8 }}}
\newcommand{\sixteenth}{ {{1 \over 16 }}}
\def\var{ \hbox{var} }
\newcommand{\HB}{ {\hbox{{\scriptsize\em HB}\/}} }
\newcommand{\MGMC}{ {\hbox{{\scriptsize\em MGMC}\/}} }
\newcommand{\gtilde}{ {\widetilde{G}} }
\newcommand{\longto}{\longrightarrow}

%\ltapprox and \gtapprox produce > and < signs with twiddle underneath
\def\spose#1{\hbox to 0pt{#1\hss}}
\def\ltapprox{\mathrel{\spose{\lower 3pt\hbox{$\mathchar"218$}}
 \raise 2.0pt\hbox{$\mathchar"13C$}}}
\def\gtapprox{\mathrel{\spose{\lower 3pt\hbox{$\mathchar"218$}}
 \raise 2.0pt\hbox{$\mathchar"13E$}}}

\newcommand{\scra}{{\cal A}}
\newcommand{\scrb}{{\cal B}}
\newcommand{\scrc}{{\cal C}}
\newcommand{\scrd}{{\cal D}}
\newcommand{\scre}{{\cal E}}
\newcommand{\scrf}{{\cal F}}
\newcommand{\scrg}{{\cal G}}
\newcommand{\scrh}{{\cal H}}
\newcommand{\scrk}{{\cal K}}
\newcommand{\scrl}{{\cal L}}
\newcommand{\scrm}{{\cal M}}
\newcommand{\scrmvec}{\vec{\cal M}}
\newcommand{\scrp}{{\cal P}}
\newcommand{\scrr}{{\cal R}}
\newcommand{\scrs}{{\cal S}}
\newcommand{\scru}{{\cal U}}

\def\bsigma{\mbox{\protect\boldmath $\sigma$}}
\renewcommand{\Re}{\mathop{\rm Re}\nolimits}
\newcommand{\tr}{\mathop{\rm tr}\nolimits}
\newcommand{\CP}{ \hbox{\it CP\/} }

\font\specialroman=msym10 scaled\magstep1  % 12-point Special Roman (caps only)
\font\sevenspecialroman=msym7              % 7-point Special Roman (caps only)
\def\zed{\hbox{\specialroman Z}}
\def\szed{\hbox{\sevenspecialroman Z}}
\def\R{\hbox{\specialroman R}}
\def\sR{\hbox{\sevenspecialroman R}}
\def\N{\hbox{\specialroman N}}
\def\C{\hbox{\specialroman C}}
\renewcommand{\emptyset}{\hbox{\specialroman ?}}

%
% Array for subscripts
%
\newenvironment{sarray}{
          \textfont0=\scriptfont0
          \scriptfont0=\scriptscriptfont0
          \textfont1=\scriptfont1
          \scriptfont1=\scriptscriptfont1
          \textfont2=\scriptfont2
          \scriptfont2=\scriptscriptfont2
          \textfont3=\scriptfont3
          \scriptfont3=\scriptscriptfont3
        \renewcommand{\arraystretch}{0.7}
        \begin{array}{l}}{\end{array}}

\newenvironment{scarray}{
          \textfont0=\scriptfont0
          \scriptfont0=\scriptscriptfont0
          \textfont1=\scriptfont1
          \scriptfont1=\scriptscriptfont1
          \textfont2=\scriptfont2
          \scriptfont2=\scriptscriptfont2
          \textfont3=\scriptfont3
          \scriptfont3=\scriptscriptfont3
        \renewcommand{\arraystretch}{0.7}
        \begin{array}{c}}{\end{array}}

\def\bgamma{\vec{\mbox{$\gamma$}}}
\def\bt{\vec{\mbox{$t$}}}
\def\bvu{\vec{\mbox{$u$}}}
\def\bvv{\vec{\mbox{$v$}}}
\def\bsig{\vec{\mbox{$\sigma$}}}
\def\bx{\mbox{\protect\boldmath $x$}}
\def\by{\mbox{\protect\boldmath $y$}}
\def\bun{\mbox{\protect\boldmath $e$}}
\def\bu{\mbox{\protect\boldmath $u$}}
\def\bv{\mbox{\protect\boldmath $v$}}
\def\bk{\mbox{\protect\boldmath $k$}}
\def\bp{\mbox{\protect\boldmath $p$}}
\newcommand{\1}{1\!\!\!\bot}
\newcommand{\ba}{\begin{eqnarray}}
\newcommand{\ea}{\end{eqnarray}}
%

%%%%%%%%%%%%%%%%%%%%%%%%%%%%%%%%%%%%%%%%%%%%%%%%%%%%%%%%%%%%%%%%%%%%%%
%%%%%%%%%%%%%%%%%%%%%%%%%%%% FIRST PAGE %%%%%%%%%%%%%%%%%%%%%%%%%%%%%%
%%%%%%%%%%%%%%%%%%%%%%%%%%%%%%%%%%%%%%%%%%%%%%%%%%%%%%%%%%%%%%%%%%%%%%

\title{ Critical Slowing-Down in  \\
         $SU(2)$ Landau Gauge-Fixing Algorithms \\
      }

\author{
  \small  Attilio Cucchieri               \\[-0.2cm]
  \small  Tereza Mendes               \\[0.2cm]
  \small\it Department of Physics     \\[-0.2cm]
  \small\it New York University       \\[-0.2cm]
  \small\it 4 Washington Place        \\[-0.2cm]
  \small\it New York, NY 10003 USA    \\[-0.2cm]
  \small Internet:  {\tt ATTILIOC@ACF2.NYU.EDU} ,
                    {\tt MENDES@MAFALDA.PHYSICS.NYU.EDU}\\[-0.2cm]
  {\protect\makebox[5in]{\quad}}  % To force authors' names to be written
                                  %   vertically, one above another.
                                  % (\author seems to put them side-by-side
                                  %   if there is room.)
   \\
}

\vspace{0.2cm}
\maketitle
\thispagestyle{empty}   % Suppress page number on front page.

\begin{abstract}
We study the problem of critical slowing-down
for gauge-fixing algorithms (Landau gauge)
in $SU(2)$ lattice gauge theory on a $2$-dimensional lattice.
We consider five such algorithms, and lattice sizes ranging from
$8^{2}$ to $36^{2}$ (up to $64^2$ in the case of Fourier
acceleration). 
We measure four different observables
and we find that for each given algorithm they all have the same
relaxation time within error bars. We obtain that: the so-called
{\em Los Alamos} method has dynamic critical exponent $z \approx 2$,
the {\em overrelaxation} method and the {\em stochastic overrelaxation}
method have $z \approx 1$, the so-called
{\em Cornell} method has $z$ slightly smaller than $1$
and the {\em Fourier acceleration} method completely eliminates critical
slowing-down. A detailed discussion and analysis of the tuning
of these algorithms is also presented.

\end{abstract}

\clearpage

%%%%%%%%%%%%%%%%%%%%%%%%%%%%%%%%%%%%%%%%%%%%%%%%%%%%%%%%%%%%%%%%%%%%%%
%%%%%%%%%%%%%%%%%%%%%%%%%%%%%%% INTRODUCTION %%%%%%%%%%%%%%%%%%%%%%%%%
%%%%%%%%%%%%%%%%%%%%%%%%%%%%%%%%%%%%%%%%%%%%%%%%%%%%%%%%%%%%%%%%%%%%%%

\section{Introduction}

The lattice formulation of QCD
provides a regularization which makes the gauge group compact,
so that the Gibbs average of any gauge-invariant quantity is
well-defined and thus gauge fixing is, in principle,
not needed. However, to better
understand the relationship between continuum and lattice models, one is
led to consider gauge-dependent quantities
on the lattice as well, which requires gauge fixing.
It is therefore important to devise numerical algorithms
to efficiently gauge fix a lattice configuration.
The efficiency of these algorithms is even more important
if the problem of existence of Gribov copies in the lattice
is taken into account \cite{MO1}--\cite{FH}.
In fact, since usually
it is not clear how an algorithm selects among different Gribov copies,
numerical results using gauge fixing
could depend\footnote{~Of course, this does not apply to
gauge-invariant quantities.} on the gauge-fixing algorithm, making
their interpretation conceptually difficult.
In these cases, in order to analyze the dependence on the Gribov
ambiguity \cite{NP,FH,PPPTV1},
several Gribov copies
of the same thermalized configuration
have to be produced, and therefore gauge fixing is extensively used.

\vskip 0.4cm

In this paper we study the problem of fixing the standard
lattice Landau gauge condition
\cite{W2,MO2}. As we will see in the next section,
this condition is formulated as a minimization problem
for the energy
of a nonlinear $\sigma$-model with disordered
couplings.
Two basic {\em deterministic} {\em local} algorithms
have been introduced 
to achieve this goal. Following \cite{SS}
we will refer\footnote{~Some
of the algorithms we consider here are also well-known
for other numerical problems, and
are usually referred to with other names.
In particular, if we consider the problem of solving
a linear system of equations \cite{Axe} (or the equivalent problem of
minimizing the related quadratic action), then the
Los Alamos method corresponds to a non-linear version
of the {\em Gauss-Seidel} method,
while the overrelaxation method
-- and the Cornell method (see Sections 5 and 7) ---
correspond to non-linear versions of the {\em successive
overrelaxation} method. On the other hand, from the point of view
of minimizing a function, the Cornell method is
a {\em steepest-descent} method \cite{PTVF}, since the function
is minimized in the direction of the local downhill gradient,
while the Los Alamos method brings the ``local'' function to its unique
absolute minimum. In both cases, the idea is to decrease the
value of the minimizing function monotonically, namely these are {\em
descent} methods \cite{O}.
} to them as the ``Los Alamos'' method
\cite{GGKPSW,FG} and the ``Cornell'' method \cite{DBKKLWRS}.
Both methods are expected to perform poorly \cite{SS},
especially as the volume of the lattice increases,
due to the phenomenon of {\em critical slowing-down}, which
afflicts Monte Carlo simulations of critical phenomena as well as some
deterministic iterative methods, such as our gauge
fixing.\footnote{~For an excellent introduction to the problem
of critical slowing-down in 
Monte Carlo simulations, and also to some deterministic examples,
see \cite{S1}.}
Roughly speaking, the problem is that,
since the updating is local, the 
``information'' travels at each step only from one lattice site 
to its nearest neighbors,
executing a kind of random walk through the lattice; as a result,
in order to get any significant change in configuration, we must wait
a time of the order of the square
of the typical ``physical length'' of the system, which is 
in our case the lattice size $N$.
More precisely, the {\em relaxation
time} (measured in sweeps)
for conventional local algorithms diverges as the square of the linear
size of the system, or equivalently these methods have
{\em dynamic critical exponent} $z=2$.
This $\,N^2\,$ behavior is not a serious difficulty for small lattices,
and other aspects of the algorithms may be of greater importance in this
case; but as one deals with progressively larger lattices
(in order to approach the continuum limit)
this factor constitutes a severe limitation.

To overcome this problem, two solutions are available:
one can either modify the local update,
in such a way that the ``length'' of the move in the configuration
space is increased \cite{MO3}, and therefore this space is explored faster,
or introduce some kind of global updating, in order
to speed up the relaxation of
the long-wavelength modes, which are the slow ones.
Various methods, based on
these two ideas, have been proposed: overrelaxation \cite{MO3},
stochastic overrelaxation \cite{FG}, multigrid schemes \cite{FG,GS,HLSS,HLS},
Fourier acceleration \cite{DBKKLWRS}, wavelet acceleration
\cite{DN}, etc.
By analogy with other deterministic problems \cite{Axe,PTVF,O}
and with Monte
Carlo simulations \cite{S2,A},
the ``improved'' local
algorithms (such as overrelaxation and stochastic overrelaxation)
are expected to reduce but not eliminate critical slowing-down ($z\approx1$,
as opposed to $z\approx2$ for conventional methods),
while global methods (such as multigrid, Fourier acceleration,
etc.)  hope
to eliminate critical slowing-down completely ({\em i.e.}, $z=0$).
In any case, a precise determination of the dynamic
critical exponents for the
different algorithms is of great importance, as are analyses
and comparisons between the methods applied to the problem at hand.
Such analyses have been partially done in some of the references given
here, but we feel that systematic studies of these methods {\em for the
specific problem of Landau gauge fixing} are lacking,
especially with 
regard to the evaluation of dynamic critical exponents, although some of the
algorithms we consider were
extensively analyzed when applied to other numerical problems.
The only algorithm thoroughly studied
in the past was the
multigrid \cite{H,FG,HLSS,HLS} and we will not consider it here.
For our research, we have decided to study,
besides the two ``basic'' algorithms (Los Alamos and Cornell),
the standard overrelaxation and the stochastic overrelaxation
(both applied to the Los Alamos method), which are very
appealing
for their simplicity and almost absence of overhead.
We will also study the
Fourier acceleration (applied to the Cornell method),
for which not so many studies have been done
up to now,
and which is claimed \cite{MMS} to be the best method available today for
Landau gauge fixing.

In this work our goals are:
\begin{enumerate}
\item studying the critical slowing-down for the various
      algorithms and finding accurate values for their dynamic 
critical exponents,
\item analyzing the relative size 
      of several quantities, usually employed in the
      literature to test the convergence of the gauge fixing, in order
      to understand which of them should be used in
      practical computations,
\item doing, when necessary, a careful tuning of the algorithms,
      checking the ``empirical'' formulae commonly used for the optimal
      choice of the parameters, or finding a simple {\em prescription}
      when this formula is not known,
\item comparing the computational costs of the algorithms,
\item doing a simple analysis of the algorithms in order to get at least an
      idea of how they deal with the problem of critical slowing-down.
\end{enumerate}
Regarding the last point, we will essentially
try to review what is known about the algorithms
under analysis. Here, in fact, we are unable to do a more rigorous analysis
(in the style of \cite{N} or \cite{Wo} for the gaussian model),
due to the non-linearity of the update and the presence of
``random'' link-dependent coupling constants.

\vskip 0.4cm

The paper is organized as follows.
In Section 2 we present a pedagogical review of the
problem of Landau gauge fixing on the lattice. In particular,
to make the analysis
of the efficiency of the local algorithms a little more quantitative, we
introduce two functions: the variation of the
minimizing function at each site $\,\Delta {\widetilde {\cal E}}(\bx)\,$,
and the ``length'' of the update $\,{\cal D(\bx)}\,$,
interpreted as a move in the configuration space $\{g(\bx)\}\,$.
In Section 3 we define the update for the various algorithms and we
explain how each one fights the problem of critical slowing-down.
The quantities for which the relaxation time $\tau$ will be measured
are introduced in Section 4.
In Section 5 the problem of the {\em tuning} of the algorithms is
addressed, and we try a quantitative analysis
to find simple formulae for the optimal
choice of the parameters.
Finally, in Section 6, we give some more details about the
numerical simulations and the computational aspects of our work
and, in Section 7, we present our results and conclusions.

\vskip 0.4cm

The main difficulty we had to overcome in this project was the
severe lack of computer time, which restricted us to
dealing with the $SU(2)$ gauge
group, instead of the more interesting $SU(3)$ case,
and with small lattice sizes.
On the other hand, the use of $SU(2)$ makes
the analysis of the algorithms simpler,
and with the values of $\beta$ that we consider (see Section 4)
no significant finite-size corrections are expected to occur and the
use of small lattices is justified.

A further difficulty is the definition of {\em constant physics},
necessary for finding the dynamic critical exponents
that characterize each algorithm (see Section 4).
This definition is very simple only in dimension $d = 2$ and this is the
case we will consider here, leaving the extension of this work to
four dimensions to a future paper \cite{future}.

Nevertheless, we believe that this work presents a comprehensive 
analysis and comparison of the different methods considered, and
enough evidence for the evaluation of their dynamic critical
exponents. 
Our findings for the exponents are basically confirmations
of what is generally accepted, with the exception of the value slightly
smaller than one for the Cornell method, a fact that we try to interpret
in Sections 5 and 7.

The total computer
time used in our simulations was
about 225 hours on an IBM RS-6000/250 machine.

%%%%%%%%%%%%%%%%%%%%%%%%%%%%%%%%%%%%%%%%%%%%%%%%%%%%%%%%%%%%%%%%%%%%%%
%%%%%%%%%%%%%%%%%%%%%% LANDAU GAUGE-FIXING CONDITION %%%%%%%%%%%%%%%%%
%%%%%%%%%%%%%%%%%%%%%%%%%%%%%%%%%%%%%%%%%%%%%%%%%%%%%%%%%%%%%%%%%%%%%%

\section{Landau Gauge-Fixing Condition on the Lattice}

Let us consider a standard Wilson action for
$SU(2)$ lattice gauge theory in $d$ dimensions \cite{W1}~:
\be
S\left(\left\{U\right\}\right) \equiv
   \frac{4 a^{d-4}}{g^{2}_{0}} \, \frac{1}{2} \sum_{\mu, \nu = 1}^{d}\,
      \sum_{\bx}\, \biggl\{ \, 1 \, - \frac{1}{2} Tr \left[\
U_{\mu}(\bx) \; U_{\nu}(\bx + a \bun_{\mu}) \;
U_{\mu}^{-1}(\bx + a \bun_{\nu}) \; U_{\nu}^{-1}(\bx) \right]
\, \biggr\}
\ee
where $U_{\mu}(\bx) \in SU(2)$ are the link variables,
$g_{0}$ is the bare coupling constant, $a$ is the lattice spacing and
$\bun_{\mu}$ is a unit vector in the positive $\mu$ direction.
Sites are labeled by $d$-dimensional vectors $\bx\,$. The lattice
size in the $\mu$ direction is $L_{\mu} \equiv a\,N_{\mu}$ where
$N_{\mu}$ is an integer. We assume periodic boundary conditions,
{\em i.e.} $\bx + L_{\mu} \bun_{\mu} \equiv \bx$,
and the lattice volume is given by
\be
V \equiv \prod_{\mu = 1}^{d} L_{\mu} \, = \, a^{d} \prod_{\mu = 1}^{d} N_{\mu}
\;\mbox{.}
\ee
The gauge field is defined as
\be
A_{\mu}(\bx) \,\equiv\, \frac{1}{2 a g_{0}}
              \left[ \; U_{\mu}(\bx) -
                    U_{\mu}^{\dagger}(\bx)
              \; \right]
\; \mbox{;}
\ee
this variable
approaches the classical gauge potential in the continuum limit.

To fix the Landau gauge we look for a local minimum\footnote{~Here
we do not consider the problem of searching for the {\em absolute}
minimum of the minimizing function $\,{\cal E}\,$, which defines 
the so-called {\em minimal Landau gauge} \cite{Z}.}
of the function
\cite{W2,MO2}
\be
{\cal E}\left(\left\{ g \right\}\right) \; \equiv \; 1 -
        \frac{a^{d}}{2\,d\,V} \sum_{\mu = 1}^{d} \sum_{\bx} \,
            \frac{1}{2}\, \mbox{Tr} \left[
        \; U_{\mu}^{\left( g \right)}(\bx) +
     U_{\mu}^{\left( g \right)\dagger}(\bx)
                \; \right] 
\ee
keeping the configuration $\left\{ U_{\mu}(\bx) \right\}$ fixed.
Here each $\,g(\bx)\in SU(2)\,$ is a site variable,
$\,{\cal G} \equiv \left\{ g(\bx) \right\}\,$ is a gauge
transformation, and $\,U_{\mu}^{\left( g \right)}(\bx)\,$
is given by
\be
U_{\mu}(\bx) \;
     \rightarrow \; U_{\mu}^{\left( g \right)}(\bx) \; \equiv \; g(\bx) \;
          U_{\mu}(\bx) \;
                   g^{\dagger}(\bx + a \bun_{\mu})
\label{eq:gaugetr}
\; \mbox{.}
\ee

\vskip 0.4cm

We use the following parametrization for the matrices
$U \in SU(2)$:
\be
U \, \equiv \, u_{0} \, \1 + i \bvu \cdot \bsig \, = \,
    \left( \begin{array}{cc}
          u_{0} + i u_{3} & u_{2} + i u_{1} \\
        - u_{2} + i u_{1} & u_{0} - i u_{3}
    \end{array} \right)
\label{eq:upar}
\ee
where $\1$ is the $2 \times 2$ identity matrix,
the components of $\bsig \equiv ( \sigma^{1} \mbox{,}
\, \sigma^{2} \mbox{,} \, \sigma^{3} ) $ are the Pauli matrices,
$u_{0} \in \R$, $\bvu \in \R^{3}$ and $ \,u_{0}^{2} + \bvu \cdot \bvu = 1$.
Therefore
\be
U^{-1} = U^{\dagger} = u_{0}\, \1 - i\,\bvu \cdot \bsig
\label{eq:uinvpar}
\ee
and from equations (\ref{eq:upar}) and (\ref{eq:uinvpar})
it follows that
\be
\mbox{Tr} \; U = \mbox{Tr} \; U^{\dagger} = 2 \, u_{0} \; \mbox{.}
\label{eq:traceu}
\ee
By using equations (\ref{eq:uinvpar}) and (\ref{eq:traceu})
we can also write
\be
U^{-1} = U^{\dagger} = \1 \mbox{Tr}\,U - U
\label{eq:udagger}
\; \mbox{.}
\ee
If $V = v_{0} + i\, \bvv \cdot \bsig$ is another $SU(2)$ matrix
then, again using (\ref{eq:upar}) and (\ref{eq:uinvpar}), we
obtain
\ba
\frac{1}{2} \mbox{Tr} \; ( \, U \; V^{\dagger} \, ) & = &
   u_{0} \, v_{0} \, + \,
        \bvu \cdot \bvv  \\
& = &  u \cdot v \; \mbox{,}
\label{eq:scalarp}
\ea
where the last step follows
if we interpret a matrix $U \in SU(2)$ as a four-dimensional unit
vector $u \equiv (u_{0} \mbox{,} \, \bvu )$.
Finally, if $U \in SU(2)$, the matrix
\be
A \equiv \frac{1}{2} \,
              [ \; U - U^{\dagger} \; ]
\ee
belongs to the $\germansu\, (2)$ Lie algebra and is parametrized as
\be
A = \,i\,\bvu \cdot \bsig =
               \left( \begin{array}{cc}
          i u_{3} & u_{2} + i u_{1} \\
          - u_{2} + i u_{1} & - i u_{3}
    \end{array} \right) \; \mbox{.}
\ee
This matrix is traceless, anti-hermitian and
\be
\mbox{Tr} \, (\, A A^{\dagger}
      \, ) =
\mbox{Tr} \, ( \, - A^{2} \, ) = 2 \; \bvu \cdot \bvu \; \mbox{.}
\ee

\vskip 0.4cm

Let us now consider a one-parameter subgroup $g(\tau; \bx )$ of $SU(2)$
defined by
\be
g(\tau; \bx) \equiv
   \exp \left[ \; \tau \, \gamma (\bx) \; \right]
\label{eq:oneparam}
\ee
where the parameter $\tau$ is a real number and the $\gamma$'s
are {\em fixed} elements of the $\germansu\, (2)$ Lie algebra given by
\be
\gamma (\bx) \, \equiv \, i\,\bgamma (\bx) \cdot \bsig
\ee
with $\bgamma (\bx) \in \R^{3}$ for all $\bx$.
Then, for fixed $\{ U_{\mu}\left(\bx\right) \}$,
the minimizing function ${\cal E}$ can be regarded as a function of the
parameter $\tau$, and its first derivative is given by the well-known
expression
\be
{\cal E}^{'}(0) \, = \, 
  \frac{a^{d+2} g_{0}}{d\,V} \, \sum_{\bx} \,
      \left[\gamma(\bx)\right]_{j}  \, 
          \left[\left(\nabla\cdot A^{\left( g \right)}\right)(\bx)\right]_{j}
  \; \mbox{,}
\label{eq:Ederiv}
\ee 
where the sum in the color index $j$ is understood $(j = 1, 2, 3)$ and
\be
\left[\left(\nabla\cdot A \right)(\bx)\right]_{j} \equiv \frac{1}{a}
  \sum_{\mu = 1}^{d} \, \left[ A_{\mu} (\bx) -
                  A_{\mu} (\bx - a \bun_{\mu}) \right]_{j}
\label{eq:diverA}
\ee
is the lattice divergence of
\be
\left[A_{\mu} (\bx) \right]_{j} \, \equiv \, \frac{1}{2\,i} \, \mbox{Tr}
                  \left[ A_{\mu}(\bx) \, \sigma^{j} \right]
    \; \mbox{.}
\ee
If $\{ U_{\mu}\left(\bx\right) \}$ is a stationary point of
${\cal E}(\tau)$ at $\tau = 0$ ({\em i.e.} $g(\tau,\bx) = \1,
\forall \bx$) then we have
${\cal E}^{'}(0) = 0$ for all  $\{ \gamma_{j} (\bx) \}$. This
implies
\be
\left[\left(\nabla \cdot A \right)(\bx)\right]_{j} = 0 \qquad \qquad
  \forall \; \; \; \bx \mbox{,} \; \; j
\label{eq:diverg0}
\ee
which is the
lattice formulation of the usual local Landau gauge-fixing condition
in the continuum.
By summing equation (\ref{eq:diverg0}) over the components $x_{\mu}$ of $\bx$
with $\mu \neq \nu$ and using the periodicity of the lattice, it is easy
to check that \cite{MO2} if the Landau gauge-fixing condition is
satisfied then the quantities
\be
Q_{\nu}(x_{\nu}) \, \equiv \, \sum_{\mu \neq \nu} \,
     \sum_{x_{\mu}} \, A_{\nu}(\bx)  \qquad \qquad
  \phantom{forall} \; \; \; \nu = 1\mbox{,}\ldots\,\mbox{,} d
\label{eq:charges}
\ee
are constant, {\em i.e.} independent of $x_{\nu}$.
From this it immediately follows that the longitudinal
gluon propagator at zero three-momentum
\ba
D_{\nu \nu}(x_{\nu}) & \equiv &
    \sum_{\mu \neq \nu} \, \sum_{x_{\mu}} \, \frac{1}{2}\,
      \mbox{Tr} \,\<\,  A_{\nu}(\bx)\, A_{\nu}(0) \, \>
\label{eq:gluonprop} \\
 & = &
\frac{1}{2}\, \mbox{Tr} \,\<\, Q_{\nu} (x_{\nu}) \,
  A_{\nu}(0) \, \>
\ea
is also constant.

\vskip 0.4cm

The numerical problem we have to solve is therefore the following:
for a given ({\em i.e.} fixed) thermalized lattice configuration 
$\{ U_{\mu}(\bx) \}$,
we want to find a gauge transformation $\{ g (\bx) \}$ 
which brings the function
\ba
{\cal E}\left(\left\{ g \right\}\right) & = &
 1 \,-\, \frac{a^{d}}{2\,d\,V} \sum_{\mu = 1}^{d} \sum_{\bx} \,
         \mbox{Tr} \; U_{\mu}^{\left( g \right)}(\bx)
\nonumber \\
&=& 1 \,-\,
\frac{a^{d}}{2\,d\,V} \sum_{\mu = 1}^{d} \sum_{\bx} \,
         \mbox{Tr} \left[ \;
g(\bx) \;
          U_{\mu}(\bx) \;
                   g^{\dagger}(\bx + a \bun_{\mu}) \;
   \right]
\label{eq:spingl}
\ea
to a local minimum, starting from a configuration $g(\bx) = \1$ for all
$\bx$.
In order to achieve this result it is
sufficient to find an iterative process which,
from one iteration step to the next, decreases the
value of the minimizing function monotonically (descent methods).
In fact, since
${\cal E}$ has a lower bound of 0 (and an upper bound of 2),
an algorithm of this kind is expected to converge.

To find a simple iterative algorithm which minimizes ${\cal E}$
one may choose to update a single site variable $g(\by)$
at a time. In this case the minimizing function ${\cal E}$
becomes
\ba
{\widetilde {\cal E}}\left[ g(\by) \right] & = & constant \,-\,
 \frac{a^{d}}{2\,d\,V} \, \mbox{Tr} \, w(\by)
\ea
where 
\be
w(\by) \;\equiv\; g(\by) \, h(\by)
\label{eq:wdefi}
\ee
and the ``single-site effective magnetic field'' $\,h(\by)\,$ is given by
\be
h(\by) \equiv \,
        \sum_{\mu = 1}^{d}
            \left[ \; U_{\mu}(\by) \;
                 g^{\dagger}(\by + a \bun_{\mu}) +
    U_{\mu}^{\dagger}(\by - a \bun_{\mu}) \;
        g^{\dagger}(\by - a \bun_{\mu}) \; \right] \; \mbox{.}
\label{eq:hdefinition}
\ee
The matrices $h(\by)$ and $w(\by)$ are proportional
to $SU(2)$ matrices, namely they can be written as
\ba
h(\by) &\equiv& \sqrt{\det h(\by)} \; {\widetilde h}(\by)
\label{def_ztilde} \\
w(\by) &\equiv& \sqrt{\det w(\by)} \;
  {\widetilde w}(\by)
\label{eq:htilda}
\ea
with ${\widetilde h}(\by), {\widetilde w}(\by) \in SU(2)$ and
\be
{\cal N}(\by)\, \equiv\,
\sqrt{\det h(\by)} \, = \, \sqrt{\det w(\by)}
   \; \mbox{.}
\label{eq:defN}
\ee
Let us also define
\be
{\cal T}(\by) \equiv \mbox{Tr}\, {\widetilde w}(\by)
\, = \, \mbox{Tr}\,w(\by) \, / \, {\cal N}(\by)
\; \mbox{.}
\label{eq:calT}
\ee

\vskip 4mm

We want to consider the single site update
$g(\by) \,\rightarrow \, g^{(new)}(\by)\,$, which can alternatively
be looked at as the multiplicative update
\be
g(\by) \,\rightarrow \, g^{(new)}(\by) \equiv R^{(update)}(\by) \, g(\by)
\; \mbox{,}
\label{eq:gmove}
\ee
with $R^{(update)}(\by) \in SU(2)$.
Under this update,
the variation of the minimizing function is given by
\ba
\Delta {\widetilde {\cal E}}(\by) & = &
 \frac{a^{d} {\cal N}(\by)}{2\,d\,V}
      \, \mbox{Tr} \, \biggl\{  \left[ \; g(\by) \, - \,
                     g^{(new)}(\by) \; \right]
              {\widetilde h}(\by)  \biggr\}
\label{eq:delta}  \\
& = & \frac{a^{d} {\cal N}(\by)}{2\,d\,V}
      \, \mbox{Tr} \, \biggl\{  \left[ \; 1 \, - \,
                     R^{(update)}(\by) \; \right]
              {\widetilde w}(\by)  \biggr\}
\label{eq:delta1}
\;\mbox{.}
\ea
To measure the length of the move $g(\by) \, \rightarrow \,
g^{(new)}(\by)$ we can use\footnote{~This choice is not
useful in the case of {\em global} gauge transformations; in fact, for 
this kind of transformations we obtain a non-zero value for ${\cal
D}(\by)$ even though they do not really represent a move in the
configuration space.} the quantity \cite{Z}
\ba
{\cal D}\left[g(\by),\, g^{(new)}(\by)\right] \equiv \,
\; {\cal D}(\by) & \equiv &
 \sqrt{\, \frac{1}{2} \, \mbox{Tr} \,\biggl\{
     \left[ \, g(\by) - g^{(new)}(\by) \,
   \right] \left[ \, g(\by) - g^{(new)}(\by) \, \right]^{\dagger}
\biggr\}
 \, } \nonumber \\[0.2 cm]
& = & \sqrt{\, 2 \, - \, \mbox{Tr} \,\left[ \, g^{(new)}(\by) \,
            g^{\dagger}(\by) \, \right] \, } \nonumber \\[0.3 cm]
& = & \sqrt{\, 2 \, - \, \mbox{Tr} \, R^{(update)}(\by)
\, }
\label{eq:delta2}
\label{eq:dismat}
\ea
which satisfies the defining properties of a {\em distance} function for
any set of matrices and, if we interpret $SU(2)$ matrices as
four-dimensional unit vectors, it coincides with the standard euclidean
distance in $\R^{4}\,$
[see formula (\ref{eq:scalarp})].

In the next section the local quantities 
$\,\Delta {\widetilde {\cal E}}(\by)\,$
and $\,{\cal D}(\by)\,$ will be used to illustrate the performance of
the different local methods considered.
In particular, their expressions will be written completely in
terms of the tuning parameter for the algorithm (if needed),
the square root $\,{\cal N}(\by)\,$ of the determinant 
of $w(\by)$ and $h(\by)$,
and the trace $\,{\cal T}(\by)\,$ of the normalized matrix
$\,{\widetilde w}(\by)\,$.

%%%%%%%%%%%%%%%%%%%%%%%%%%%%%%%%%%%%%%%%%%%%%%%%%%%%%%%%%%%%%%%%%%%%%%
%%%%%%%%%%%%%%%%%%%%%%%%%%%% THE ALGORITHMS %%%%%%%%%%%%%%%%%%%%%%%%%%
%%%%%%%%%%%%%%%%%%%%%%%%%%%%%%%%%%%%%%%%%%%%%%%%%%%%%%%%%%%%%%%%%%%%%%

\section{The Algorithms}

In this section we will describe the five algorithms for which
we want to analyze the critical slowing-down.
In particular, we will compare the implementation and performance
of the four local algorithms\footnote{~A simple comparison of this kind is not
possible for the Fourier acceleration method.}
considered in Sections 
\ref{section:los_alamos}--\ref{section:stoc_overrelax},
by comparing their expressions for the quantities
$\,\Delta {\widetilde {\cal E}}(\by) \,$ and
$\,{\cal D}(\by)\,$ introduced in the previous section.

As explained before, $\,\Delta {\widetilde {\cal E}}(\by)\,$
measures
by how much the ``local energy'' is reduced in a
single step of the algorithm at site $\by$, while
$\,{\cal D}(\by)\,$ measures by how much the
configuration at site $\by$ was effectively changed.
Therefore they
represent the two (possibly conflicting) tasks that
a local algorithm is expected to perform: minimizing the energy at
every site and
at the same time moving efficiently through the configuration space.

As we will see in Section \ref{section:los_alamos} below, the Los Alamos
method has the ``best'' possible value for 
$\,\Delta {\widetilde {\cal E}}(\by) \,$, {\em i.e.} it brings the
``single site energy'' to its absolute minimum in one iteration.
We take the Los
Alamos method as a basis for comparing all the other local algorithms
we consider, which will typically have smaller 
$\,\Delta {\widetilde {\cal E}}(\by) \,$ (in magnitude) and larger
$\,{\cal D}(\by)\,$ than the Los Alamos method, and will perform
better.

In order to make this comparison more quantitative
we will also look at the ratios
\be
{\cal R}_{{\cal E}}(\by) \equiv
\frac{
\Delta {\widetilde {\cal E}}^{(LosAl.)}(\by)
\, - \, \Delta {
\widetilde {\cal E}}(\by) }{
 \Delta {\widetilde {\cal E}}^{(LosAl.)}(\by)}
\label{eq:rplot2}
\ee
and
\be
{\cal R}_{{\cal D}}(\by) \equiv
\frac{     
{\cal D}(\by)\, - \,
 {\cal D}^{(LosAl.)}(\by) }{ {\cal D}^{(LosAl.)}(\by)}
\label{eq:rplot1}
\ee
for the various methods.
They measure, respectively, the relative ``loss'' in minimizing ${\widetilde
{\cal E}}(\by)\,$ and the relative ``gain'' in the length of the update, when
compared with the Los Alamos method. In the
next subsections these two quantities will be evaluated,
for each local algorithm, as
functions of $\,{\cal N}(\by)\,$ and $\,{\cal T}(\by)\,$
(defined in the previous section)
and the tuning parameter. In
particular, their limits as $\,{\cal T}(\by)\,\to 2$ will be
computed. In fact, as
discussed in Section \ref{section:Tuning}, these limits
are useful to point out analogies between the algorithms
and compare their efficiencies in fighting critical slowing-down.

%%%%%%%%%%%%%%%%%%%%%%%%%%%%%%%%%%%%%%%%%%%%%%%%%%%%%%%%%%%%%%%%%%%%%%
%%%%%%%%%%%%%%%%%%%%%%%%%%%% LOS ALAMOS METHOD %%%%%%%%%%%%%%%%%%%%%%%
%%%%%%%%%%%%%%%%%%%%%%%%%%%%%%%%%%%%%%%%%%%%%%%%%%%%%%%%%%%%%%%%%%%%%%

\subsection{Los Alamos Method}
\label{section:los_alamos}

It is easy to see from equations \reff{eq:delta} and \reff{eq:delta1}
that the choice \cite{GGKPSW,FG}
\be
g^{(new)}(\by)\, =\, g^{(LosAl.)}(\by) \equiv {\widetilde h}^{\dagger}(\by)
\; \mbox{,}
\label{eq:update}
\ee
or
\be
R^{(update)}(\by)\, \equiv \, {\widetilde w}^{\dagger}(\by)
\label{eq:update1}
\; \mbox{,}
\ee
gives the maximum negative variation of ${\widetilde {\cal E}}$:
\be
\Delta {\widetilde {\cal E}}^{(LosAl.)}(\by) \, = \,
   \frac{a^{d} {\cal N}(\by)}{2\,d\,V} \,
      \left[ \,{\cal T}(\by) - 2 \,
     \right] \leq 0 \; \mbox {,}
\label{eq:deltamin}
\ee
where $\,{\cal T}(\by)\,$ was defined in equation \reff{eq:calT}.
In other
words, the move from
$g(\by)$ to $g^{(LosAl.)}(\by)$ brings
the function ${\widetilde {\cal E}}\left[ g(\by) \right]$
to its unique absolute minimum. For this update we have,
from equation (\ref{eq:delta2}),
\be
{\cal D}^{(LosAl.)}(\by) \, = \,
  \sqrt{ 2\, -\, {\cal T}(\by)}  \; \mbox{.}
\label{eq:dismin}
\ee

%%%%%%%%%%%%%%%%%%%%%%%%%%%%%%%%%%%%%%%%%%%%%%%%%%%%%%%%%%%%%%%%%%%%%%
%%%%%%%%%%%%%%%%%%%%%%%%%% CORNELL METHOD %%%%%%%%%%%%%%%%%%%%%%%%%%%%
%%%%%%%%%%%%%%%%%%%%%%%%%%%%%%%%%%%%%%%%%%%%%%%%%%%%%%%%%%%%%%%%%%%%%%

\subsection{Cornell Method}
\label{section:cornell}

Another possible choice for $\,R^{(update)}(\by)\,$ comes
from considering an update of the form (\ref{eq:oneparam}) with
$\tau = \alpha$ and
$\gamma(\by) = - a^{2} g_{0}
\left(\nabla\cdot A^{\left( g \right)}\right)(\by)$. Then, from
equation (\ref{eq:Ederiv}) we obtain
\be
\Delta{\widetilde {\cal E}}\left[g(\by)\right] \, = \, - \alpha
\frac{a^{d+4} g^{2}_{0}}{d\,V} \, \sum_{j = 1}^{3} \,
   \left[\left(\nabla\cdot A^{\left( g
\right)}\right)(\by)\right]^{2}_{j}
\; + \, {\cal O}( \alpha^{2} )
\ee
and it is clear that the minimizing function decreases
if $\alpha$ is small and positive. So we can define \cite{DBKKLWRS}
\be
R^{(update)}(\by) \, \equiv \, \exp\left[ - \alpha\,
  a^{2}\, g_{0}\, \left( \nabla \cdot A^{\left( g \right)} \right) (\by)
  \right]
\label{eq:cornupdate}
\; .
\ee
Since we expect
$\,a^{2}\, g_{0}\, \left( \nabla \cdot A^{\left( g \right)} \right)
(\by)\,$ to go to zero as the number of iterations increases,
we can expand $\,R^{(update)}(\by)\,$ to first order obtaining
\be
R^{(update)}(\by) \, \propto \, \left[ \1 - \alpha \, 
 a^{2}\, g_{0}\, \left( \nabla \cdot A^{\left( g \right)} \right) (\by)
\right]
\; \mbox{;}
\label{eq:gCornell}
\ee
here $\propto$ indicates that the matrix on the l.h.s. is proportional
to the one on the r.h.s. (namely it has to be reunitarized) and
the parameter $\alpha$ is a {\em positive real number} which has to be
properly tuned, depending on the considered configuration, as discussed
below.

If we notice that the matrix $w(\by)$ [defined in \reff{eq:wdefi}]
satisfies the relation
\be
w(\by) = w^{\dagger}(\by) \, + \,
         2\, a^{2}\, g_{0} \,
         \left( \nabla \cdot A^{\left( g \right)} \right) (\by)
\label{eq:hA}
\ee
we can rewrite equation (\ref{eq:gCornell}) as
\be
R^{(update)}(\by) \, \propto \, \1 + \frac{\alpha}{2} \,
    \left[ \, w^{\dagger}(\by) \, - w(\by) \, \right]
\ee
and, by using equation\footnote{~Note that equation \reff{eq:udagger}
holds also for {\em multiples} of $SU(2)$ matrices 
such as $ w^{\dagger}(\by)\,$.}
(\ref{eq:udagger}) with $U = w^{\dagger}(\by)\,$, we obtain
\be
R^{(update)}(\by) \, \propto \, \left[ 1 \, - \, \frac{\alpha}{2} \,
   \mbox{Tr} \, w^{\dagger}(\by) \, \right]
  \, \1 \, + \, \alpha \, w^{\dagger}(\by) 
\; \mbox{.}
\ee
Finally,
reunitarizing $R^{(update)}(\by)$
and using (\ref{eq:gmove}) we have
\be
g^{(new)}(\by) \, = \, g^{(cornell)}(\by) \,  \equiv
\, \frac{
\alpha \, {\cal N}(\by) \, {\widetilde w}^{\dagger}(\by) \, + \,
\left[ 1 \, - \, \alpha\, {\cal N}(\by) \, {\cal T}(\by)  / 2\, \right]
 \, \1 }{
     \sqrt{\, 1 + \alpha^2\, {\cal N}^{2}(\by)  \left[
            \, 1 - {\cal T}^{2}(\by) / 4 \, \right]
        }} \, g(\by)
\; \mbox{,}
\label{eq:movecorn}
\ee
where ${\widetilde w}(\by)$, ${\cal N}(\by)$ and
${\cal T}(\by)$ are defined respectively in (\ref{eq:htilda}),
\reff{eq:defN} and (\ref{eq:calT}). In this case the
variation of the minimizing function is given by
\be
\Delta {\widetilde {\cal E}}^{(cornell)}(\by)
 = \frac{a^{d} {\cal N}(\by)}{2\,d\,V}
      \, \left\{ \; {\cal T}(\by) \, - \,
\frac{
2\, \alpha \, {\cal N}(\by) \, + \,
\left[ 1 \, - \, \alpha\, {\cal N}(\by) \, {\cal T}(\by)  / 2\, \right]
 \, {\cal T}(\by) }{
     \sqrt{\, 1 + \alpha^2\, {\cal N}^{2}(\by)  \left[
            \, 1 - {\cal T}^{2}(\by) / 4 \, \right]
        }}
              \;  \right\} \; \mbox{.}
\ee
Since ${\cal T}(\by)$ is in the interval $[-2,\,2]$ and
$\alpha$ is {\em positive} this quantity is negative or zero\footnote{~To
see this notice that $\Delta{\widetilde {\cal E}}$ is zero at the end points
${\cal T}(\by) = \pm 2$ and negative for ${\cal T}(\by) = 0$, that
there are no other zeros in this interval iff
$\alpha\,{\cal N}(\by) \in (0,\,2]\,$, and that for $\alpha\,{\cal N}(\by) >
2$ the variation $\Delta {\widetilde {\cal E}}$ approaches zero from
above as ${\cal T}(\by)$ goes to two.}
iff $\alpha\,{\cal N}(\by) \in (0,\,2]$.
Therefore the algorithm converges only if $\alpha$ is positive and small
enough.
On the other hand, if we evaluate the length of the move
$g(\by) \, \rightarrow \, g^{(cornell)}(\by)$ we obtain
from equation (\ref{eq:delta2})
\be
{\cal D}^{(cornell)}(\by) \, = \, \sqrt{2} \,
\left( 1 - \left\{
         1 + \alpha^2 \, {\cal N}^{2}(\by)  \left[
\, 1 - \frac{{\cal T}^{2}(\by)}{4} \, \right] \right\}^{-1/2} \,
  \right)^{1/2}
\; \mbox{;}
\ee
namely $\alpha$ should not be too small otherwise
this length goes to zero.

It is easy to check that in the limit of
$\,{\cal T}(\by)\to 2\,$ we get for the ratios
$\,{\cal R}_{{\cal E}}(\by)\,$ and 
$\,{\cal R}_{{\cal D}}(\by)\,$ [defined respectively in
\reff{eq:rplot2} and \reff{eq:rplot1}] 
\be
{\cal R}^{(Cornell)}_{{\cal E}}(\by) \, \rightarrow
  \, \left[ \, \alpha\,{\cal N}(\by) \, - \, 1 \,\right]^{2}
\ee
and
\be
{\cal R}^{(Cornell)}_{{\cal D}}(\by) \, \rightarrow \,
\left[ \, \alpha\,{\cal N}(\by) \, - \, 1 \, \right]
\; \mbox{.}
\ee
Therefore, as $\,{\cal T}(\by)\,$ approaches $2$
we have that, at least
for the case $\,1<\alpha\,{\cal N}(\by)<2\,$,
the gain in the length of the move with respect to the Los Alamos
method is linear in $\,\left[ \, \alpha\,{\cal N}(\by) \, - \, 1 \, \right]\,$,
while the loss in the minimizing of the energy is quadratically small.
This illustrates why the algorithm will perform better than the Los Alamos
method. For further discussion, see Section 5.

%%%%%%%%%%%%%%%%%%%%%%%%%%%%%%%%%%%%%%%%%%%%%%%%%%%%%%%%%%%%%%%%%%%%%%
%%%%%%%%%%%%%%%%%%%%% OVERRELAXATION METHOD %%%%%%%%%%%%%%%%%%%%%%%%%%
%%%%%%%%%%%%%%%%%%%%%%%%%%%%%%%%%%%%%%%%%%%%%%%%%%%%%%%%%%%%%%%%%%%%%%

\subsection{Overrelaxation Method}
\label{section:overr}

The standard overrelaxation method \cite{MO3} is a local
algorithm in which, instead of using the update
\be
g(\by) \,
     \rightarrow \, g^{(LosAl.)}(\by) = {\widetilde w}^{\dagger}(\by)
       \, g(\by)
\ee
described in Section 3.1,
we use the substitution
\be
g(\by) \,
     \rightarrow \, g^{(new)}(\by)\, =\, g^{(over)}(\by)
    \equiv \left[\, {\widetilde w}^{\dagger}(\by) \,
           \right]^{\omega}
       \, g(\by) \; \mbox{,}
\ee
where the overrelaxation parameter $\omega$ varies in
the interval $\,1 < \omega < 2\,$
and has an optimal value which is volume- and problem-dependent.
Of course, for $\omega = 1$, we have $g^{(over)}(\by) = g^{(LosAl.)}(\by)$
while, for $\omega = 2$,
we obtain
\be
g^{(over)} (\by)\, = \, {\widetilde w}^{\dagger}(\by) \,
      {\widetilde w}^{\dagger}(\by) \, g(\by) \; \mbox{,}
\ee
and it is easy to check that [see equation (\ref{eq:delta1})]
\be
\Delta {\widetilde {\cal E}}(\by)
       \; = \; \frac{a^{d} {\cal N}(\by)}{2\,d\,V}
      \, \mbox{Tr} \, \left[ \; {\widetilde w}(\by) \, - \,
         {\widetilde w}^{\dagger}(\by) \; \right] \; = \; 0 \;
   \mbox{,}
\label{eq:nomove}
\ee
namely
for $\omega = 2$
the algorithm does not converge, as the energy
is never decreased.

Finally,
for $1 < \omega < 2$, we can write
\be
g^{(over)}(\by) = \left[\, {\widetilde w}^{\dagger}(\by) \,
           \right]^{\omega - 1} g^{(LosAl.)}(\by) \; \mbox{.}
\ee
Therefore we can interpret this update as a move from
$g(\by)$ to $g^{(over)}(\by)$ ``passing through'' $g^{(LosAl.)}(\by)$.
In this way, the minimizing function ${\widetilde {\cal E}}
\left[ g(\by) \right]$
will not go to its absolute minimum, even though
its variation $\Delta {\widetilde {\cal E}}$ will still be
negative.
For computing $\left[ \, {\widetilde w}^{\dagger}(\by)
\, \right]^{\omega}$ one uses
the binomial
expansion\footnote{~We could also write
the matrix $\,{\widetilde w}(\by)\,$ as
\begin{equation}
{\widetilde w}(\by)\, =\, \1 \,\cos \gamma \,+\,i\,
              \vec{n} \cdot \bsig \, \sin \gamma \, = \,
                    \exp \left(\, i\, \gamma\,
        \vec{n}
\cdot \bsig\,\right)
\end{equation}
with $\,\gamma \in [-\pi, \pi)\,$,
$\vec{n} \in \R^{3}$ and $\vec{n}\cdot\vec{n} \,=\,1$. Then
$\left[ \, {\widetilde w}^{\dagger}(\by)\, \right]^{\omega}$
would be given by $\exp \left(\,- i\, \gamma\,\omega\,
\vec{n} \cdot \bsig\,\right)$ where the
product $\,\gamma\,\omega\,$ should be considered modulo $\,2\,\pi\,$
so that $\,\gamma\,\omega\,\in [-\pi, \pi)\,$.
In any case we are interested in the limit in which
${\widetilde w}^{\dagger}(\by)$ approaches the identity matrix $\1$, namely
$\gamma\to 0\,$. By expanding $\exp \left(\,- i\, \gamma\,\omega\,
\vec{n} \cdot \bsig\,\right)$ around
$\gamma = 0\,$, and reunitarizing
we obtain again formula \reff{eq:overupdate}.
}
\be
\left[ \, {\widetilde w}^{\dagger}(\by)\, \right]^{\omega} \, = \,
    \sum_{n=0}^{\infty} \, \frac{ \Gamma(\omega + 1)}{
         n! \; \Gamma(\omega + 1 - n)} \,
        \left[\, {\widetilde w}^{\dagger}(\by) - \1\, \right]^{n}
\; \mbox{.}
\ee
Since the matrix ${\widetilde w}^{\dagger}(\by)$ is expected to
converge to the identity matrix $\1$, this series
can be truncated after a
few terms, followed by
reunitarization of the resulting matrix; for example, if only two terms are
kept, we have
\be
g^{(over)}(\by) \, = \,  \frac{ 
\biggl\{ \1 +
     \omega \left[\, {\widetilde w}^{\dagger}(\by) - \1\, \right]
       \, \biggr\} }{
     \sqrt{ 1 + \omega  \left( \omega - 1\right)
                   \left[\,2 - {\cal T}(\by)
             \,\right]
        }} \, g(\by)
\; \mbox{.}
\label{eq:overupdate}
\ee
In this way the
variation of the minimizing function is given by
[ see equation (\ref{eq:delta1}) ]
\be
\Delta {\widetilde {\cal E}}^{(over)}(\by)
 \, =\,  \frac{a^{d} {\cal N}(\by)}{2\,d\,V}
      \, \left\{ \; {\cal T}(\by) \, - \,
           \frac{\left(1 - \omega \right) {\cal T}(\by) \, +
   \, 2 \, \omega }{
      \sqrt{ 1 + \omega  \left( \omega - 1\right)
                   \left[\,2 - {\cal T}(\by)
             \,\right]
        }}
              \;  \right\} \; \mbox{.}
\ee
Since $\, {\cal T}(\by) \in [-2,2]$ and $\omega \in (1,2)$ it is easy
to check that this variation is always negative or
zero\footnote{~Furthermore, it can be
proved that, if $\omega > 1/2\,$,
the variation $\Delta{\widetilde {\cal E}}$ is zero only
at ${\cal T}(\by) = 2\,$ while, if $\omega < 1/2\,$, this happens at
both end points ${\cal T}(\by) = \pm 2\,$.
Then it is easy to check
that $\Delta{\widetilde {\cal E}}$ is
always negative or zero if $\omega \in [0,2]\,$.
}.
The length of the move [formula (\ref{eq:dismat})] is given
by
\be
{\cal D}^{(over)}(\by) \, = \,
 \left\{\, 2 -
\frac{ 2 -  \, \omega \left[\,2\,-\,{\cal T}(\by) \,\right]
      }{ \sqrt{ 1 + \omega  \left( \omega - 1\right)
                   \left[\,2 - {\cal T}(\by)
             \,\right]
        } } \right\}^{1/2}
\;\mbox{.}
\ee

As an illustration of the improvement with respect to the
Los Alamos method, let us consider $\omega$ slightly
larger than $1$. Expanding the expressions
for $\,\Delta {\widetilde {\cal E}}^{(over)}\,$ and
$\,{\cal D}^{(over)}\,$ around $\,\omega = 1$, we obtain
\be
\Delta {\widetilde {\cal E}}^{(over)}(\by)
 \, = \, \frac{a^{d} {\cal N}(\by)}{2\,d\,V}
      \, \left[ \, {\cal T}(\by) \, - \, 2 \, \right] \,
\left\{ \, 1
    \, - \frac{1}{4} \left(\omega  - 1\right)^{2} \,
         \left[ \,2 + {\cal T}(\by) \,\right] \,
  + \, {\cal O}\left( \left(\omega  - 1\right)^{3} \right) \,
       \right\}
\ee
\vskip 2mm
and
\be
{\cal D}^{(over)}(\by) \, = \,
  \sqrt{\, 2 -  {\cal T}(\by)\, } \,
   \left\{\, 1 + \frac{1}{4} \left( \omega - 1\right) \,
       \left[\, 2 + {\cal T}(\by)\, \right] \, +
 \, {\cal O}\left( \left(\omega  - 1\right)^{2} \right) \,
           \right\} 
\; \mbox{,}
\ee
\vskip 2mm
{\noindent 
namely the correction with respect to the variation (\ref{eq:deltamin})
is positive and quadratic in $(\omega  - 1)$,
while the correction with respect to the
length (\ref{eq:dismin}) is positive and linear in
$(\omega  - 1)$.}
Thus, already for a value of 
$\omega$ slightly larger than one,
what we lose in minimizing
${\widetilde {\cal E}}$, compared with the Los Alamos method,
is ``smaller'' than what we gain in the length
${\cal D}(\by)$ of the move for the update and therefore the
relaxation process should be speeded up.

More generally, these features can be seen
from the behavior of the
quantities $\,{\cal R}^{(over)}_{{\cal E}}(\by)\,$ and
$\,{\cal R}^{(over)}_{{\cal D}}(\by)\,$ [defined respectively in
\reff{eq:rplot2} and \reff{eq:rplot1}].
As an example, we plot
in Figure 1 these two ratios
as a function of ${\cal T}(\by)$ and with $\, \omega = 1.9\,$.
In particular, in the limit ${\cal T}(\by) \rightarrow 2$ we obtain
\be
{\cal R}^{(over)}_{{\cal E}}(\by) \, \rightarrow
  \, \left( \, \omega \, - \, 1 \,\right)^{2}
\label{eq:r1overr}
\ee
and
\be
{\cal R}^{(over)}_{{\cal D}}(\by) \, \rightarrow \,
\left( \,\omega \, - \, 1\, \right)
\; \mbox{.}
\label{eq:r2overr}
\ee
It is interesting to note that the behavior is
qualitatively the same as the one
for the Cornell method, discussed at the end of the previous subsection.

\vskip 0.5cm

\begin{figure}
\epsfxsize=0.4\textwidth
\centerline{\epsffile{fig1.ps}}
\caption{~Plot
of the ratios $\,{\cal R}^{(over)}_{{\cal E}}(\by)\,$ and
$\,{\cal R}^{(over)}_{{\cal D}}(\by)\,$ as functions of
$\,\mbox{Tr}\,{\widetilde w}={\cal T}\,$,
 for comparison between the
overrelaxation method at $\,\omega=1.9\,$ and the Los Alamos method. 
}
\end{figure}

%%%%%%%%%%%%%%%%%%%%%%%%%%%%%%%%%%%%%%%%%%%%%%%%%%%%%%%%%%%%%%%%%%%%%%
%%%%%%%%%%% STOCHASTIC OVERRELAXATION METHOD %%%%%%%%%%%%%%%%%%%%%%%%%
%%%%%%%%%%%%%%%%%%%%%%%%%%%%%%%%%%%%%%%%%%%%%%%%%%%%%%%%%%%%%%%%%%%%%%

\subsection{Stochastic Overrelaxation Method}
\label{section:stoc_overrelax}

The stochastic overrelaxation method \cite{FG}
is also a local
algorithm. In this case, instead of always applying the 
descent step $g(\by) \rightarrow g^{(LosAl.)}(\by)$ one uses
the new update
\be
g(\by) \, \rightarrow \, g^{(new)}(\by) \, = \,
      g^{(stoc)}(\by) \, \equiv \,
   \left\{ \begin{array}{ll}
        \left[\, {\widetilde w}^{\dagger}(\by) \, \right]^{2} \, g(\by) &
                   \quad \mbox{with probability $p$} \\
      \phantom{ } & \phantom{ } \\
          g^{(LosAl.)}(\by) &
                   \quad \mbox{with probability $1 - p$}
        \end{array} \right.
\label{eq:stocupdate}
\ee
with $0 < p < 1$.
Of course for $p = 0$ this algorithm coincides with the Los Alamos
method while, for $p = 1$, it does not converge at all
since, as we saw in formula (\ref{eq:nomove}), the
value of the minimizing function ${\widetilde {\cal E}}
\left[ g(\by) \right]$ remains constant.
However, for $p \in (0,\,1)$,
the fact that, with probability $p$, a big move is done
in the configuration space without changing the value of
${\widetilde {\cal E}}\left[ g(\by) \right]$
has, again, the capability of speeding up the
relaxation process. To check this point we can compute
the length of the move
\be
g(\by) \, \rightarrow \,
\left[\, {\widetilde w}^{\dagger}(\by) \, \right]^{2} \, g(\by)
\; \mbox{;}
\label{eq:stochupdatep}
\ee
by using equation (\ref{eq:udagger}) we can rewrite
$\left[\, {\widetilde w}^{\dagger}(\by) \, \right]^{2}$ as
\be
\left[\, {\widetilde w}^{\dagger}(\by) \, \right]^{2} \, = \,
  {\widetilde w}^{\dagger}(\by) \, {\cal T}(\by) \, - \, \1
\label{eq:overhead}
\ee
and [see equation (\ref{eq:delta2})] we
easily obtain\footnote{~It is also straightforward to check that
for this algorithm, as $\,{\cal T}(\by)\,$
goes to two, the ratios
$\,{\cal R}_{{\cal E}}(\by)\,$ and
$\,{\cal R}_{{\cal D}}(\by)\,$, defined at the beginning of Section 3,
are both equal to one, with probability $\,p\,$, and to
zero, with probability $1 - p$.}
\be
{\cal D}^{(stoc)}(\by) \, = \,
    \sqrt{\, 4 \, - \, {\cal T}^{2}(\by) }
\, = \, \sqrt{\, 2 \, + \, {\cal T}(\by) } \;\,
{\cal D}^{(LosAl.)}(\by)
\; \mbox{.}
\ee

Roughly speaking, we can say that the stochastic overrelaxation
method alternates updates that give the maximum negative variation
of $\,{\widetilde {\cal E}}\,$ with steps that produce ``very long''
moves in the configuration space, without increasing the value
of the minimizing function. This is similar in spirit to
the idea behind the so-called {\em hybrid} version of
overrelaxed algorithms \cite{A,Wo,BW}, which are
used to speed up Monte Carlo simulations with
spin models, lattice gauge theory, etc. In these algorithms,
$n$ microcanonical (or energy conserving) update sweeps
are done followed by one standard local ergodic update
(like Metropolis or heat-bath) sweeping over the lattice.
Actually, for the gaussian model, it has been proven
\cite{Wo} that the best result is obtained when the $n$ microcanonical
steps are picked at random, namely when $n$ is the average number of
microcanonical sweeps between two subsequent ergodic updates. This is
essentially what is done in the stochastic overrelaxation method,
with $n / (n + 1)$ equals in average to $p$ or, equivalently,
$n$ equals in average to $p / (1 - p)$.

Finally, formula (\ref{eq:overhead}) can also be used in order to
reduce the overhead of the algorithm; we can in fact rewrite
(\ref{eq:stocupdate}) as
\be
      g^{(stoc)}(\by) \, = \,
   \left\{ \begin{array}{ll}
          {\widetilde h}^{\dagger}(\by) \, {\cal T}(\by) \,
                        - \, g(\by) &
                   \quad \mbox{with probability $p$} \\
      \phantom{ } & \phantom{ } \\
          {\widetilde h}^{\dagger}(\by) &
                   \quad \mbox{with probability $1 - p$}
        \end{array} \right.
\ee
[ where $\,{\widetilde h}\,$ was introduced in \reff{def_ztilde} ],
namely instead of computing matrix products of the form
${\widetilde h}^{\dagger}(\by)\, g^{\dagger}(\by)\, 
{\widetilde h}^{\dagger}(\by)$
we just have to do a simple linear combination of
${\widetilde h}^{\dagger}(\by)$ and $g(\by)\,$.

%%%%%%%%%%%%%%%%%%%%%%%%%%%%%%%%%%%%%%%%%%%%%%%%%%%%%%%%%%%%%%%%%%%%%%
%%%%%%%%%%%%%%%%%%%%% FOURIER ACCELERATION %%%%%%%%%%%%%%%%%%%%%%%%%%%
%%%%%%%%%%%%%%%%%%%%%%%%%%%%%%%%%%%%%%%%%%%%%%%%%%%%%%%%%%%%%%%%%%%%%%

\subsection{Fourier Acceleration}
\label{sec:Fourier}

The idea of Fourier acceleration \cite{DBKKLWRS} is very
simple. If we consider the Cornell method, it is immediate from
formula (\ref{eq:cornupdate}) that
its convergence is controlled by the quantity
$a^{2}\,g_{0}\,(\nabla\cdot A^{\left( g \right)})(\by)\,$.
For the {\em abelian case in the continuum}
it can be shown \cite{DBKKLWRS}, by analyzing the relaxation of the
different components of this matrix in momentum space,
that
\be
(\nabla\cdot {\widetilde A}^{\left( g \right)})_{t}\,(\bk) \,
\approx \, 
(\nabla\cdot {\widetilde A}^{\left( g \right)})_{0}\,(\bk) \,
\exp\left(- \alpha\,p^{2}(\bk)\,t\,\right)
\; \mbox{,}
\label{eq:decayFFT}
\ee
namely each component decays 
as $\,\exp ( - \alpha\,p^{2}(\bk)\,t\,)\,$, where $\,t\,$
indicates the number of sweeps. This means that their
decay rates are approximatively equal to
$1 / (\alpha\, p^{2}(\bk)\,)\,$. Therefore, if we choose
$\,\alpha \propto p^{- 2}_{max}\,$
we obtain that the
slowest mode has a relaxation time\footnote{~For an
exact definition see formula (\ref{eq:taudef}).} $\tau$ proportional to
$p^{2}_{max} / p^{2}_{min}\,$. In a lattice with $N$ points on each
side we have $p^{2}_{max} \propto {\cal O}(1)$ and
$p^{2}_{min}  \propto {\cal O}( N^{-2} )\,$, namely
\be
\tau \propto N^{2} \; \mbox{.}
\label{tauN2}
\ee
From this analysis it is clear how, for the abelian case, the
relaxation process can be speeded up:
given the matrix
$a^{2}\,g_{0}\,(\nabla\cdot A^{\left( g \right)})(\bx)\,$, we
take its Fourier transform, we multiply each component in momentum
space by $p^{2}_{max} / p^{2}(\bk)\,$, we evaluate the
inverse Fourier transform and, finally, the result is used in equation
(\ref{eq:gCornell}) instead of the original matrix
$a^{2}\,g_{0}\,(\nabla\cdot A^{\left( g \right)})(\by)\,$.
In a more concise form we can write:
\be
R^{(update)}(\by) \, \propto \, \1 - {\widehat F}^{-1}\left\{ 
         \, \frac{p^{2}_{max}}{p^{2}(\bk)} \,
                {\widehat F} \left[ \, \alpha \,
 a^{2}\, g_{0}\, \left( \nabla \cdot A^{\left( g \right)} \right) 
\right]\, \right\}(\by)
\; \mbox{;}
\label{eq:fftupdate}
\ee
where ${\widehat F}$ indicates the Fourier transform
and ${\widehat F}^{-1}$ is its inverse.
In this way we should obtain that the components in momentum space
of $a^{2}\,g_{0}\,(\nabla\cdot A^{\left( g \right)})(\by)\,$ decay
as
\be
\exp \left(  - \alpha\,p^{2}(\bk)\,t \, \frac{p^{2}_{max}}{p^{2}(\bk)} \right)
  \, = \, \exp \left(  - \alpha\,p^{2}_{max}\,t\right)
\ee
which, with the choice $\alpha \propto p^{- 2}_{max}\,$, gives
$\tau \propto{\cal O}(1)$ for every component.

\vskip 0.4cm

Of course, for the non-abelian case, this analysis becomes
more complicated.
Nevertheless, it is still believed\footnote{~Note that the
behavior \reff{tauN2} corresponds to dynamic critical exponent $2$
for the Cornell method. This is
in contradiction with the analysis given in Section 5, which 
predicts an exponent $z\approx 1$, by analogy with the overrelaxation
method. Our results (see Section 7) corroborate the latter prediction.
}
that the Cornell method will have \cite{SS,DBKKLWRS} the behavior 
\reff{tauN2}, and that the strategy to be used in the Fourier
acceleration is given by the modified update \reff{eq:fftupdate}.
The main difficulty arises from the fact (see again \cite{DBKKLWRS})
that, instead of the eigenvalues of the laplacian $\partial^{2}$
({\em i.e.} instead of $p^{2}(\bk)\,$), we have to consider
the eigenvalues of the operator $\partial\cdot D$, where $D$ is the
covariant derivative. Thus, the relaxation time $\tau$ will
be proportional to the ratio of the largest over the
smallest eigenvalue of $\partial\cdot D\,$ and the
eigenvectors of this operator should be used to
decompose the divergence of $A^{\left( g \right)}$. This is
of course not easy to be implemented in a numerical
simulation and, therefore, the eigenvectors of
the laplacian are used also in the non-abelian case.
The hope is that the non-abelian nature of the fields does not make
the behavior of $\partial\cdot D\,$, in momentum space,
too different from that of the laplacian. Actually this is
more than just hope since, in the lattice Landau gauge, the
link variables $\left\{U_{\mu}(\bx)\right\}$
are fixed as close as possible to
the identity matrix (see \cite{Z}, Appendix A) and therefore the operator
$\partial\cdot D\,$ should be, in some sense, a ``small modification'' of the
laplacian.

\vskip 0.4cm

The practical implementation of the Fourier acceleration
is also quite simple. In fact, we have to evaluate
$a^{2}\,g_{0}\,(\nabla\cdot A^{\left( g \right)})(\bx)$
at each lattice site and then use formula (\ref{eq:fftupdate}) ---
where now ${\widehat F}$ has to be interpreted as
a standard Fast Fourier Transform
subroutine \cite{PTVF} --- to find
$R^{(update)}(\by)$ at the given lattice site.
Of course, to reduce the number of times the FFT 
is used, a checkerboard update should be employed.
For our
FFT subroutine we used as a 
basis in momentum space the functions 
$\,\exp(\,2\,\pi\,i\,\bk\cdot\bx\,)\,$,
where $\bk$ has components $k_{\mu}$ given by
\be
a\,k_{\mu}\,N_{\mu}\, =\, 0\mbox{,}\,1\mbox{,}\,\ldots \mbox{,}\,
                         N_{\mu} - 1
\ee
and $\,\mu = 1, \,
\dots, d\,$. In this case the eigenvalues of the laplacian operator
$\partial^{2}$
are given by the well-known formula
\be
\bp^{2} \, \equiv\,  \frac{ 4}{a^2} \, \sum_{\mu = 1}^{d} \,
   \sin^{2}\left( \, \pi \,a\,k_{\mu} \, \right)
\ee
and the largest eigenvalue $\,\bp^{2}_{max}\,$ is obtained
when
\be
a\,k_{\mu}\,N_{\mu}\,=\,\lfloor\frac{N_{\mu}}{2}\rfloor
\ee
for all $\,\mu = 1, \,\dots, d\,$.

Finally, it is important to
observe that formula \reff{eq:fftupdate} is singular when
$\,p^{2}(\bk)\,$ is zero. However, the zero-frequency mode of
the divergence of $A^{\left( g \right)}$ does not contribute
to the update (\ref{eq:gCornell}); in fact, by using the periodicity
of the lattice and formula \reff{eq:diverA}, it is easy to check that
\be
a^{2}\,g_{0}\,\sum_{\bx}\,
   (\nabla\cdot A^{\left( g \right)})(\bx)\,=\, 0
\; \mbox{.}
\ee
Thus, in equation (\ref{eq:fftupdate}) when $\,p^{2}(\bk)\,$
is equal to zero, we can set the
value of $p^{2}_{max} / p^{2}(\bk)\,$
to any finite value without affecting
the performance of the method.

%%%%%%%%%%%%%%%%%%%%%%%%%%%%%%%%%%%%%%%%%%%%%%%%%%%%%%%%%%%%%%%%%%%%%%
%%%%%%%%%%%%%%%%%%%% CRITICAL SLOWING-DOWN %%%%%%%%%%%%%%%%%%%%%%%%%%%
%%%%%%%%%%%%%%%%%%%%%%%%%%%%%%%%%%%%%%%%%%%%%%%%%%%%%%%%%%%%%%%%%%%%%%

\section{Critical Slowing-Down}
\label{section:CSD}

To check the convergence of the gauge fixing
\cite{SS,DBKKLWRS,MO3,HLSS,PPPTV2} several quantities
have been introduced:
\ba
e_{1}(t) & \equiv & {\cal E}(t - 1) - {\cal E}(t) \\[0.25cm]
e_{2}(t) & \equiv & \frac{a^{d+4}\, g^{2}_{0}}{V} \sum_{\bx} \,
\sum_{j = 1}^{3} \, \Big[ 
    \left( \nabla \cdot A \right) (\bx) \Big]_{j}^{2}
\label{e2}
        \\
e_{3}(t) & \equiv & \frac{a^{d}}{V} \sum_{\bx} \,
    \frac{1}{2} \, \mbox{Tr} \,
   \left\{ \left[\, \1 - R^{(update)}(\bx) \, \right]
      \left[\, \1 -  R^{(update)}(\bx) \, \right]^{\dagger}
     \right\}
   \\[0.1cm]
e_{4}(t) & \equiv & \max_{\bx} \left[ \, 1 \, - \, \frac{1}{2} \,
   \mbox{Tr}\,R^{(update)}(\bx) \right] 
\label{e4} \\[0.2cm]
e_{5}(t) & \equiv & 1\, -\, \frac{a^{d}}{2\,V} \,
       \sum_{\bx} \, \mbox{Tr}\,R^{(update)}(\bx)
\ea
where $t$ indicates the number of lattice sweeps and,
when not indicated, 
the expressions on the r.h.s are evaluated after $t$ sweeps
of the lattice are completed.
All these
quantities are expected to converge to zero
exponentially
and with
the same rate \cite{SS} even though their sizes can differ considerably.
Actually, it is easy to see that
\be
e_{3} \, = \, \frac{2 \, a^{d}}{V} \sum_{\bx} \,
\left[ \, 1 \, - \, \frac{ \mbox{Tr}\,R^{(update)}(\bx) }{2} \, \right]
\, = \, 2\, e_{5} \, \approx\,2\,e_{4}
\; \mbox{.}
\ee
By using equation (\ref{eq:hA}) we can also rewrite $e_{2}$ as
\be
e_{2} \, = \, \frac{a^{d}}{V} \sum_{\bx} \, {\cal N}^{2}(\bx)\,
\, \left[ \, 1 \, - \, \frac{ {\cal T}^{2}(\bx) }{4} \, \right]
\; \mbox{.}
\label{eq:e2second}
\ee
We know that, if the algorithm converges, the matrix
$R^{(update)}(\bx)$ should approach
the identity matrix $\1$ as the number of iteration
increases or, equivalently, that its trace
should be very close to two at large $\,t\,$.
This implies
[ see the expressions of $\,R^{(update)}(\bx)\,$ for the various
local methods: formulae (\ref{eq:update1}), (\ref{eq:movecorn}),
(\ref{eq:overupdate}) and (\ref{eq:stocupdate}) ] that
$\,{\widetilde w}^{\dagger}(\bx)\to\1\,$, and therefore
${\cal T}(\bx)$ is also very close to two\footnote{~From
equations (\ref{eq:delta1}) and (\ref{eq:delta2}) it is clear that
also $\Delta {\widetilde {\cal E}}(\bx)$
and ${\cal D}(\bx)$ go to zero.
In particular, since $e_{4}$ is proportional to $\max_{\bx}
\left[ {\cal D}(\bx)\right]^{2}\,$,
it is obvious that ${\cal D}(\bx)$ goes to zero exponentially.
This tells us that, when the condition ${\cal T}(\bx)\protect\ltapprox 2$
is satisfied (usually after a few sweeps), the algorithm
``moves'' very slowly
through the configuration space and therefore
improving the accuracy of the gauge fixing becomes very costly.}.
Therefore if $e_{4}$ is of order $\epsilon \ll 1$ then also
$e_{3}\,$, $e_{5}$ and $e_{2}$ should be of this order;
taking into account these relations, we decided to look only at the
quantities $e_{1}$, $e_{2}$ and $e_{4}\,$.

We also expect that the spatial fluctuations
of the quantities $Q_{\nu}(x_{\nu})$, defined in (\ref{eq:charges}),
and of the longitudinal gluon propagators, defined
in (\ref{eq:gluonprop}),
go to zero exponentially. Indeed, the fact 
that $Q_{\nu}(x_{\nu})$ should be constant is being increasingly
used as a check of the accuracy of the gauge fixing \cite{MMS,BPS}.
To check this more precisely, we introduced a new quantity
defined as
\be
e_{6}(t) \, \equiv \, \frac{1}{d} \,
  \sum_{\nu = 1}^{d} \, \frac{1}{3\,N_{\nu}}
\sum_{j = 1}^{3} \, \sum_{x_{\nu} = 1}^{N_{\nu}} \, \frac{
    \left[ \,  Q_{\nu}(x_{\nu}) - {\widehat Q}_{\nu}  \,
      \right]_{j}^{2}}{ \left[ {\widehat Q}_{\nu} \right]_{j}^{2} }
\label{eq:e6}
\ee
where
\be
{\widehat Q}_{\nu} \,
\equiv \, \frac{1}{N_{\nu}} \, \sum_{x_{\nu} = 1}^{N_{\nu}} \,
            Q_{\nu}(x_{\nu}) 
\; \mbox{.}
\ee

For each of these quantities,
in the limit of large $\,t\,$,
we can introduce \cite{H}
a relaxation time $\tau_{i}$ by the relation
\be
e_{i}(t) \, \approx \, c_{i} \,\exp\left(\, - \, t / \tau_{i} \, \right)
\; \mbox{,}
\label{eq:expei}
\ee
namely
\be
\tau_{i} \, \equiv \, \lim_{t\to\infty} \, \frac{- \, 1}{
             \ln\left[\,e_{i}(t+1) / e_{i}(t)\,\right]}
\label{eq:taudef}
\; \mbox{.}
\ee
As said before, we expect all these relaxation times to coincide and
be equal to $\tau$.

\vskip 0.4cm

To analyze the critical-slowing down of an algorithm we have to
measure $\,\tau\,$ for different pairs\footnote{~From now on
we always consider lattices with $N_{\mu} = N$ for all $\mu = 1, \,
\dots, d\,$.} of $\,N\,$ and $\, \beta = 4 / g_{0}^{2}\,$,
but at ``constant
physics''. This means that we have to keep the ratio $\,N / \xi\,$
constant, where the correlation length $\,\xi\,$ is given
by the inverse of the
square root of the {\em string tension} $\,\kappa\,$, {\em i.e.} $\,\xi
= 1 / \sqrt{\kappa}\,$. 
The string tension for two-dimensional $SU(2)$ lattice gauge theory 
(in the spin-$\half$ representation) is given, in the {\em infinite volume }
limit, by \cite{DM} 
\be
\kappa\, = \, - \log \frac{ I_{2}(\beta) }{ I_{1}(\beta) }
\; \mbox{,}
\ee
where $\,I_{n}\,$ is the modified Bessel function. Thus we have
\be
\xi\,= \, \frac{1}{\sqrt{- \log \frac{ I_{2}(\beta) }{ I_{1}(\beta) }}}
\label{eq:csi}
\ee
which, in the limit of large $\beta\,$, gives
\be
\xi\,=
\,\sqrt{\frac{2\, \beta}{3}} \, \left[ \, 1 \, + \,
                 \frac{1}{4\,\beta} \, + \,
                       {\cal O}(\beta^{-2}) \, \right]
\; \mbox{.}
\ee
Therefore a constant ratio $\,N / \xi\,$ is equivalent, in this limit,
to keeping the ratio $\,N^{2} / \beta\,$ fixed.
The values for the pairs $(N, \beta)$
have been chosen
in such a way that $\,N \gg \xi\,$;
thus the finite size effects should be negligible.
All the pairs $(N, \beta)$ used are reported\footnote{~For
the case of the Fourier acceleration
we considered $N = 8\mbox{,}\,16\mbox{,}\,
32\mbox{,}\,64$. Note that we restricted our lattice sizes
to powers of $2$, because of the way in which the
Fast Fourier Transform subroutine
we used \cite{PTVF} is designed. (The application of this routine
is not limited to these lattice sizes, and it can be modified to work
with general values of $\,N\,$, but the use of powers of $2$
makes it most efficient.)
}
in Table \ref{Table.Nb}. In the same table we report the value of
the corresponding correlation length $\,\xi\,$ obtained by using
equation \reff{eq:csi}. We have chosen $\,N^{2} / \beta\,=\,32\,$
and $\,N / \xi\,\approx\,7$.

\vskip 4mm

Once all these values of $\tau$
are obtained, we can try a fit of
the form
\be
\tau \, = \, c\, N^{z}
\label{eq:taufit}
\ee
and find the dynamic critical exponent $z$ for that algorithm. It is
important to recall that the value of $z$ obtained in this
way is independent of the 
``constant physics'' chosen ($N^{2} / \beta = 32$ in our case).
On the contrary, this is not the case for the constant $\,c$. In particular
we expect the relaxation time $\tau$, and therefore $\,c$,
to {\em increase} as the
link couplings $\,U_{\mu}(\bx)\,$ in \reff{eq:spingl}
become more ``random'', {\em i.e} as 
$\beta$ decreases for a given lattice size $N$ (and the value
of the ratio $N^{2} / \beta$ increases).

\begin{table}
\addtolength{\tabcolsep}{-1.0mm}
\hspace*{-1.0cm}
\protect\footnotesize
\begin{center}
\begin{tabular}{|| c || c | c | c | c | c | c | c | c | c ||}
\hline
\hline
$ N $ & $ 8 $ & $ 12 $ & $ 16 $ & $ 20 $ & $ 24 $
     & $ 28 $ & $ 32 $ & $ 36 $ & $ 64 $ \\ \hline
$ \beta $ & $ 2.0 $ & $ 4.5 $ & $ 8.0 $ & $ 12.5 $ & $ 18.0 $ & $ 24.5 $
       & $ 32.0 $ & $ 40.5 $ & $ 128.0 $ \\ \hline
$ \xi $ & $ 1.09 $ & $ 1.65 $ & $ 2.24 $ & $ 2.83 $
       & $ 3.42 $ & $ 4.00 $ & $ 4.58 $ & $ 5.16 $ & $ 9.22 $ \\ \hline
\hline
\end{tabular}
\end{center}
\caption{~The pairs $(N, \beta)$ used for the simulations and
          the correlation length $\xi$ evaluated in the infinite
          volume limit.}
\label{Table.Nb}
\vspace*{0.3cm}
\end{table}

%%%%%%%%%%%%%%%%%%%%%%%%%%%%%%%%%%%%%%%%%%%%%%%%%%%%%%%%%%%%%%%%%%%%%%
%%%%%%%%%%%%%%%%%%%%%%%%%% TUNING %%%%%%%%%%%%%%%%%%%%%%%%%%%%%%%%%%%%
%%%%%%%%%%%%%%%%%%%%%%%%%%%%%%%%%%%%%%%%%%%%%%%%%%%%%%%%%%%%%%%%%%%%%%

\section{Tuning of the Algorithms}
\label{section:Tuning}

\begin{table}
\addtolength{\tabcolsep}{-1.0mm}
\hspace*{-1.0cm}
\protect\footnotesize
\begin{center}
\begin{tabular}{|| l |  c | c ||}
\hline
\hline
algorithm & $ a(\by) $ & $ b(\by) $ \\
\hline
\hline
Los Alamos & $ 1 $ & $ 0 $ \\ \hline
Cornell & $ \alpha\, {\cal N}(\by) $ &
     $ \alpha\, {\cal N}(\by) \, {\cal T}(\by) / 2 \, - \, 1 $ \\ \hline
overrelaxation & $\omega $ & $ \omega - 1 $ \\ \hline
stochastic overr. & $
 \begin{array}{ll}
          {\cal T}(\by) & \quad \mbox{with probability $p$} \\
             1 & \quad \mbox{with probability $1 - p$}
        \end{array} $ & $
\begin{array}{ll}
          1 & \quad \mbox{with probability $p$} \\
             0 & \quad \mbox{with probability $1 - p$}
        \end{array} $
 \\ \hline
\hline
\end{tabular}
\end{center}
\caption{~The coefficients $a(\by)$ and $b(\by) $ for the four
          local algorithms considered in this paper.}
\label{Table.ab}
\end{table}

The implementation of all the algorithms considered in this work
--- except for the Los Alamos method --- requires the tuning of a parameter:
$\alpha$ for the Cornell method and the Fourier
acceleration, $\omega$ for the overrelaxation method and
$p$ for the stochastic overrelaxation. This is, of course, a potential
disadvantage of all these methods and makes the study
of their critical slowing-down more difficult: in fact,
for each
pair $(N, \beta)\,$, the value of the parameters should be
tuned in such a way that the value of $\tau$ is minimized.
This is usually done heuristically since, as explained in the
Introduction, no rigorous analyses are available for these
algorithms. However, analytic estimates for the optimal choice
of $\omega$ are indeed known for the overrelaxation method applied
to other numerical problems \cite{Axe,PTVF,O,N,A2}; in all these cases, in
the limit of large lattice size $N$, it has been found that
\be
\omega_{opt}\, = \, \frac{2}{ 1 \, + \, C_{opt} / N}
\label{eq:omegafit}
\ee
where the constant $C_{opt}$ is problem dependent.
This result is usually adopted as a guess \cite{MO3,HLS}
also for the optimal choice of $\omega$ when the overrelaxation
method is applied to Landau gauge fixing.

In order to obtain simple formulae like (\ref{eq:omegafit})
for the optimal choice of $\alpha$ (in the
case of the Cornell method) and $p$, we decided to
compare the four local algorithms
considered in this work.
It is interesting to notice that they can
all be defined by the update\footnote{~This is, of course, not
surprising if we notice that equation \reff{eq:linearupdate} is the
most general {\em linear} {\em local}
update we can introduce to minimize the function
$\Delta {\widetilde {\cal E}}^{(over)}(\by)$.}
\be
g(\by) \, \rightarrow \, g^{(new)} \, \propto \,
            a(\by) \, {\widetilde h}^{\dagger}(\by) \, - \, b(\by)
                            \, g(\by)
\label{eq:linearupdate}
\ee
where the coefficients $a(\by)$ and $b(\by)$ are given in Table 
\ref{Table.ab}.
Moreover, we see from our simulations that in all methods,
usually after a few sweeps, we have
${\cal T}(\by) \protect\ltapprox 2 \,$. This is
evident by looking at the decay of $\,e_{2}(t)\,$
(see Figures \ref{fig:decays} and \ref{fig:decays2} for typical examples)
and considering formula \reff{eq:e2second}. Actually, since the
gauge-fixing procedure is stopped when $\,e_{2}(t)\,$ is smaller than
$10^{-12}$ (see end of Section \ref{sec:num.sim}), the condition
$\,{\cal T}\protect\ltapprox 2\,$
is satisfyed for the most part of our simulations.
With this in mind we can write
\be
g(\by) \, \rightarrow \, g^{(new)} \, \approx \,
           {\tilde a}(\by) \, {\widetilde h}^{\dagger}(\by) \, - \,
                   {\tilde b}(\by)
                            \, g(\by) 
\ee
where the coefficients ${\tilde a}(\by)$ and ${\tilde b}(\by)$
are obtained from the coefficients $a(\by)$ and $b(\by)$ by imposing
the condition ${\cal T}(\by) = 2$.

This simple analysis seems to
suggest (see Table 2) that the Cornell method is equivalent to
the overrelaxation method if we make the substitution
\be
\alpha\, {\cal N}(\by) \, \rightarrow \, \omega
\label{eq:omegainCorn}
\; \mbox{.}
\ee
The same substitution is suggested by considering
the ratios
${\cal R}_{{\cal E}}(\by)$ and ${\cal R}_{{\cal D}}(\by) $
defined in \reff{eq:rplot2} and \reff{eq:rplot1};
in fact, as ${\cal T}(\by)$ goes to $2$, we have
that the formulae for these ratios for the Cornell method and for
the overrelaxation method {\em coincide} if the above substitution is
employed (see end of Sections \ref{section:cornell} and 
\ref{section:overr} respectively).

If this analysis is correct we should obtain
for the Cornell method the same dynamic critical exponent 
as for the overrelaxation method, {\em i.e.} 
$\,z\approx 1\,$. This will be verified in Section 7.
As a further
check of this ``equivalence'' between the two
methods we can compare the optimal choices for their
tuning parameters, obtained in
our simulations. We notice that while $\,\omega\,$
and $\,\alpha\,$ are fixed parameters throughout the run, $\,{\cal N}(\by)\,$ 
changes with the iterations, and we are interested in its
value as $\,{\cal T}(\by)\to 2\,$. Moreover, 
$\,{\cal N}(\by)\,$ is a local quantity, and therefore we consider 
its space average, which can be easily estimated
in this limit. In order to do this, let us
rewrite the minimizing function as
\ba
{\cal E} & = & 
1 - \frac{a^{d}}{4\,d\,V} \sum_{\mu = 1}^{d} \sum_{\bx} \,
         \mbox{Tr} \; \left\{
   \, U_{\mu}^{\left( g \right)}(\bx) \, + \,
     \left[
U_{\mu}^{\left( g \right)}(\bx - a \bun_{\mu})\right]^{\dagger}
 \, \right\} \nonumber \\
& = & 1 -
\, \frac{a^{d}}{4\,d\,V} \sum_{\bx} \,
         \mbox{Tr} \, w(\bx) \,=\,
\, \frac{a^{d}}{4\,d\,V} \sum_{\bx} \,
         {\cal N}(\bx) \, {\cal T}(\bx)
\ea
which, in the limit ${\cal T}(\by) \rightarrow 2$, gives
\be
{\cal E} \, = \, 1 -
\, \frac{a^{d}}{2\,d\,V} \sum_{\bx} \,
         {\cal N}(\bx)
\;\mbox{.}
\ee
In other words, the space average of $\,{\cal N}(\bx)\,$ is given by
\be
 2\,d\,\left(\, 1 - {\cal E}_{stat} \, \right)
\; \mbox{,}
\label{eq:Nstima}
\ee
where $\,{\cal E}_{stat}\,$ is the value of the minimizing function
at the stationary point. Of course this value is not known
a priori but, for fixed values of $\beta$ and $V$,
its order of magnitude can be easily estimated with just a few
numerical tests.
Using this result, we are able to make a numerical comparison between the
tuning parameters for the two methods (see Section 7), finding
very good agreement.

\vskip 4mm
One may also attempt to establish a relation between the overrelaxation
and the stochastic overrelaxation. For example, we can try to write
the update (\ref{eq:stocupdate}) as an average of the two cases
with weights
$p$ and $1 - p$ obtaining (in the limit ${\cal T}(\by) \rightarrow 2$)
\be
g^{(stoc)}(\by) \, \approx \, \left(\,1 + p\right) \,
         {\widetilde h}^{\dagger}(\by) \, - \, p \, g(\by)
\;\mbox{,}
\label{eq:newstochupd}
\ee
which suggests the substitution
\be
p \, \rightarrow \, \left( \, \omega \, - \, 1\, \right)
\label{eq:omegainp}
\; \mbox{.}
\ee
However, if we now look at the ratios
${\cal R}_{{\cal E}}(\by)$ and ${\cal R}_{{\cal D}}(\by) $ for the
update \reff{eq:newstochupd} we obtain, as ${\cal T}(\by)$ goes to $2$,
\be
{\cal R}^{(stoc)}_{{\cal E}}(\by) \, \rightarrow \, p
\label{Rtop}
\ee
and
\be
{\cal R}^{(stoc)}_{{\cal D}}(\by) \, \rightarrow \, p
\; \mbox{;}
\label{Rtop2}
\ee
the second formula, if compared to \reff{eq:r2overr}, seems to
be consistent with the substitution \reff{eq:omegainp}
while the first [compared to \reff{eq:r1overr}] suggests the relation 
\be
p \, \rightarrow \, \left( \, \omega \, - \, 1\, \right)^{2}
\label{eq:omegainp2}
\; \mbox{.}
\ee
These two possibilities will be tested numerically (see
Section \ref{section:result}) by plotting the quantities
$\,( \, \omega \, - \, 1\,)\, /\, p\,$ and
$\,( \, \omega \, - \, 1\,)^{2} /\, p\,$ as a function of $\,N$.

\vskip 4mm
Finally, we do not hazard here any hypothesis on the tuning of
$\,\alpha\,$ for the Fourier acceleration method.

%%%%%%%%%%%%%%%%%%%%%%%%%%%%%%%%%%%%%%%%%%%%%%%%%%%%%%%%%%%%%%%%%%%%%%
%%%%%%%%%%%%%%%%%%%%%% NUMERICAL SIMULATIONS %%%%%%%%%%%%%%%%%%%%%%%%%
%%%%%%%%%%%%%%%%%%%%%%%%%%%%%%%%%%%%%%%%%%%%%%%%%%%%%%%%%%%%%%%%%%%%%%

\section{Numerical Simulations}
\label{sec:num.sim}

To thermalize the gauge configuration $\{ U_{\mu}(\bx) \}$
at a fixed value of the coupling $\beta\,$, we used a
standard heat-bath algorithm \cite{Cr}. In order to
optimize the efficiency of the code, we used two
different $SU(2)$ generators
(methods 1 and 2 described in Appendix A in \cite{EFGS}, with
$h_{cutoff} = 2$).

With the pairs $(N, \beta)$ that we considered (see Section
\ref{section:CSD}), we should always have $\, N \gg \xi\,$
and therefore we expect all the temporal correlations
to decay exponentially. As a check we measured, for
all the pairs $(N, \beta)$,
the {\em integrated autocorrelation time}\footnote{~See \cite{S1}
for a definition of integrated autocorrelation time and for
a description of the automatic windowing procedure used to measure
it.}
for the Wilson loop $\,W(1,1)\,$ 
and for the Polyakov loop $P\,$ (indicated respectively as
$\,\tau_{int,W_{1}}\,$ and $\,\tau_{int,P}\,$). Moreover, for the
pairs $(N, \beta)$
used for the Fourier acceleration method, we also
measured the integrated autocorrelation time $\,\tau_{int,W_{l}}\,$
for the Wilson loops $\,W(l,l)\,$ with
$\,l\,=\,2\mbox{,}\,4\mbox{,}\,8\mbox{,}\,\ldots\mbox{,}\,
N / 2\,$.

In practice, we started all our runs with a random 
$\{ U_{\mu}(\bx) \}$ configuration and we did $\,5000\,$ sweeps
for thermalization. The configurations used for gauge fixing
were separated by $\,100\,$ sweeps, in order to get
a statistically independent sample. After discarding
$\,4900\,$ sweeps out of a total of 54900 sweeps,
we evaluated $\,\tau_{int,W_{l}}\,$ and $\,\tau_{int,P}\,$
by using a window of $\,4\,\tau_{int}\,$,
which is a reasonable choice if the decay is 
roughly exponential. For the observables we considered
we obtained\footnote{~From our data it is clear that, for a
fixed lattice size,
the Wilson loop with size $l \approx \xi$ has
the largest integrated autocorrelation time among the
quantities we considered. Nevertheless, we obtain
$\,\tau_{int} \protect\ltapprox 2.5\,$ for {\em all}
our lattice sizes, except for $64^2$ (used only for
Fourier acceleration), where we get $\,\tau_{int,W_{l}}
\protect\ltapprox 10\,$ for $l \,=\,8\mbox{,}\,16$.}
$\,0.5 \protect\ltapprox \tau_{int} \protect\ltapprox 10 \,$.
Noticing
that $\tau_{int} = 0.5\,$ indicates that the data are uncorrelated,
we can conclude, as
expected, that the system decorrelates
rather fast and that the
configurations used for testing the gauge-fixing algorithms
are essentially statistically independent.

\vskip 0.4cm

The tuning of the parameters $\omega$, $p$ and $\alpha\,$ --- respectively
for the overrelaxation, the stochastic overrelaxation,
the Cornell and the Fourier acceleration methods --- was done
very carefully. More precisely, we divided it in three parts.
In the first
step, we considered a few values of the parameter spread in a
large interval. For example we used $\omega = 1.1\mbox{,}\,
1.2\mbox{,}\,\ldots\mbox{,}\,1.9$ or $p = 0.1\mbox{,}\,
0.2\mbox{,}\,\ldots\mbox{,}\,0.9\,$. For each of these values
we gauge fixed $10$ different configurations, measured all the 
$\,\tau_{i}\,$'s and averaged the results. In this way we were able to select
a smaller interval for the parameter (usually of length $\approx 0.2$ for the
overrelaxation or the stochastic overrelaxation methods)
which
was used in the second step of the tuning. In this case $20$
configurations were analyzed for each value of the parameter
(and these values were typically separated by $0.01\,$ for the
overrelaxation and stochastic overrelaxation methods).
In the last step,
the length of the interval was further reduced and $100$
configurations were gauge-fixed for each value of the parameter.

A total of $\,201.7\,$ hours of CPU were used for the four methods
requiring tuning. Of these, $\,21.6\,$ were used
for the first level of tuning, $\,36.1\,$ for the second level
and the remaining $\,144\,$ for the third level.

For the Los Alamos method no tuning is needed, but we found
that, in this case, the fluctuations for
the relaxation times are larger and therefore more
configurations ($500\,$, to be exact) had to be considered,
for a total of $\,23.3\,$ hours of CPU.

\vskip 0.4cm

Finally, for measuring the relaxation time $\,\tau_{i}\,$ (with
$\,i\,=\,1\mbox{,}\,2\mbox{,}\,4\,$ and $6$)
we did a chi-squared fitting of the functions
$\,\log e_{i}(t)\,$ which, if relation \reff{eq:expei} is
satisfied, should be a straight line. Indeed this was usually the case
already after a few initial sweeps of the lattice,
at least for the quantities $\,e_{1}\,$, $\,e_{2}\,$ and $\,e_{4}\,$.
On the contrary, the decay of $\,e_{6}\,$ is really smooth
and monotonic only for the Fourier acceleration method.
As an example we show, in Figure \ref{fig:decays},
the behavior of $\,e_2\,$ and $\,e_6\,$ as functions of $\,t\,$
for the Cornell and the Fourier acceleration method.
We also show, in Figure \ref{fig:decays2}, the decays of the four quantities
for the stochastic overrelaxation method.

We use the condition
\be
e_{2} \,\leq \,10^{-12}
\label{eq:stop}
\ee
to stop the gauge fixing, 
in order to ensure that
enough data are produced for the fitting and that, when the
procedure is stopped, essentially only the slowest mode
has survived.
To get rid of the initial fluctuations, we used only
the last $100$ data
when the total number of sweeps $N_{sw}$ was larger than $200\,$,
or the second half of the data when less than $200\,$
sweeps were necessary to fix the gauge.
For $\,i\,=\,1\,$ we have also to take into account
possible fluctuations of $\,e_{1}(t)\,$ around zero, which
appear when the minimizing function is fixed
within the machine precision. Therefore, for this quantity,
we also discarded the last $50$ sweeps, if $N_{sw} > 200$, or the
last one quarter of the data if $N_{sw}$ was smaller than
$200$.

\begin{figure}[p]
\vspace*{0cm} \hspace*{-0cm}
\begin{center}
\epsfxsize = 4in
\leavevmode\epsffile{decay_cor.ps} \\
\vspace*{5mm}
\epsfxsize = 4in
\leavevmode\epsffile{decay_fft.ps}
\end{center}
\vspace*{-5mm}
\caption{~Plot
of the decays of the quantities $\,e_2\,$ and $\,e_6\,$ as
functions of $\,t\,$ for
(a) the Cornell method, with $\alpha=0.481$, and
(b) the Fourier acceleration method, with $\alpha=0.160$.
Both plots were done for $\,16^2\,$ lattices.
}
\label{fig:decays}
\end{figure}

\begin{figure}[p]
\vspace*{0cm} \hspace*{-0cm}
\begin{center}
\epsfxsize = 4in
\leavevmode\epsffile{decay_sto.ps} \\
\end{center}
\vspace*{-5mm}
\caption{~Plot
of the decays of the quantities $\,e_1\,$, $\,e_2\,$,
$\,e_4\,$ and $\,e_6\,$ as
functions of $\,t\,$ for
the stochastic overrelaxation method, with $p = 0.75 $
on a $\,16^2\,$ lattice. The two almost superposed curves
are $\,e_2(t)\,$ and $\,e_4(t)\,$ ($\,e_2(t)\,$ is the ``smoother'' curve).
}
\label{fig:decays2}
\end{figure}

%%%%%%%%%%%%%%%%%%%%%%%%%%%%%%%%%%%%%%%%%%%%%%%%%%%%%%%%%%%%%%%%%%%%%%
%%%%%%%%%%%%%%%%%%%%%% RESULTS AND CONCLUSIONS %%%%%%%%%%%%%%%%%%%%%%%
%%%%%%%%%%%%%%%%%%%%%%%%%%%%%%%%%%%%%%%%%%%%%%%%%%%%%%%%%%%%%%%%%%%%%%

\section{Results and Conclusions}
\label{section:result}

Our final data for the relaxation times are reported,
for the different methods, in Tables \ref{Table.LosAlamos}--\ref{Table.FFT}. 
We show for each algorithm only the relaxation time $\,\tau_2\,$ for
the quantity $\,e_2\,$ defined in \reff{e2}. Indeed, we checked
that, for all methods and pairs $\,(N,\beta)\,$, the four measured relaxation
times were in agreement within error bars.
We also show the optimal choice for the tuning parameter (when needed),
the number of sweeps necessary to reach the stopping condition \reff{eq:stop},
and, for the Cornell method, the value of the minimizing function (used 
in Section 7.3 for comparison with the overrelaxation method).
Averages are taken over the different configurations that were gauge fixed
for each pair $\,(N,\beta)\,$.

%%%%%%%%%%%%%%%%%%%%%%%%%%%%%%%%%%%%%%%%%%%%%%%%%%%%%%%%%%%%%%%%%%%%%%
%%%%%%%%%%%%%%%%%%%%%%%% CRITICAL EXPONENTS %%%%%%%%%%%%%%%%%%%%%%%%%%
%%%%%%%%%%%%%%%%%%%%%%%%%%%%%%%%%%%%%%%%%%%%%%%%%%%%%%%%%%%%%%%%%%%%%%

\subsection{Critical Exponents}

We now proceed to the evaluation of the dynamic critical exponents $\,z\,$.
In Tables \ref{zfit_losal}--\ref{zfit_fft}
we present the results of
the fits to the ansatz \reff{eq:taufit} for $\,\tau_2\,$ for the
various methods. We do a weighted least-squares fit in several
``steps'', discarding at each step the values of $\,N\,$ smaller than
$\,N_{min}\,$. In this way we try to rule out some of
the smaller values of $\,N\,$ as finite-size corrections;
this is very important since we are dealing with very small
lattice sizes.
We do so for all possible values of $\,N_{min}\,$, and decide which
one gives the best fit for $z$
by comparing $\,\chi^2\,$ and confidence levels for the
different $\,N_{min}\,$'s.
As can be seen from our tables, these
finite-size
corrections are negligible already at lattice sizes $12$ or $16$.
In Figure \ref{fig:dataz}
we plot together the values of $\,\tau_2\,$ and the fitting curve for
our preferred fit for the various algorithms.

Our results for the dynamic critical exponents are in agreement with what is
generally expected, namely we find:
$\,z\approx\,2\,$ for the Los Alamos method,
$\,z\approx\,1\,$ for the overrelaxation \cite{HLS}
and the stochastic overrelaxation method, and
$\,z\approx\,0\,$ for the
Fourier acceleration method.
For the Cornell method, as mentioned in Section \ref{sec:Fourier},
a simple analysis (based on the abelian case in the
continuum) would give $\,z\approx2\,$,
and this is what is generally believed \cite{SS,DBKKLWRS}. On the 
other hand, our comparative analysis between the  Cornell
and the overrelaxation methods in Section 5
would suggest $\,z\approx1\,$.
As can be seen from
Table \ref{zfit_cor},
our results show
the latter behavior. Furthermore, in Section 7.3 we will
verify the relation between the tuning parameters for these
two methods found in Section 5.

Actually, the dynamic critical exponent $\,z\,$ for the Cornell method
is even slightly smaller than one. The good performance
of this method could be understood by noticing that
the value of $\alpha\,{\cal N}(\bx)$ changes during the
gauge-fixing process. In particular, with a few numerical
tests, it can be checked
that the space average value of ${\cal N}(\bx)$
increases with the iterations. Moreover, as we will see
below, the final value of 
$\alpha\,{\cal N}(\bx)$ is equal to the
optimal value found for $\omega$ in the overrelaxation method.
So, in a sense, we have an overrelaxation method
whose parameter $\omega \leftrightarrow \alpha\,{\cal N}(\bx)$
increases with the iterations\footnote{~It should be stressed that
this scenario is very qualitative.
In particular, as explained in
Section \ref{section:Tuning}, the relation
between $\omega\,$ and $\alpha\,{\cal N}(\bx)$ is established
only when ${\cal T}\protect\ltapprox 2$, {\em i.e.}
it should not be used for the initial sweeps of the lattice.}
from an initial
value $\omega_{0}$ up to the asymptotic value $\omega_{opt}\,$.
It is well known \cite{PTVF,A} that in overrelaxed algorithms
the optimal strategy is precisely to vary the parameter $\omega$
from an initial value $1$ to a larger asymptotic value
$\omega_{opt}\,$, which is usually
done by using Chebyshev polynomials. It is conceivable
that the Cornell method does this variation ``automatically''
and this could explain why it performs slightly better
than the overrelaxation method.

%%%%%%%%%%%%%%%%%%%%%%%%%%%%%%%%%%%%%%%%%%%%%%%%%%%%%%%%%%%%%%%%%%%%%%
%%%%%%%%%%%%%%%%%%%% CHECKING THE GAUGE-FIXING %%%%%%%%%%%%%%%%%%%%%%%
%%%%%%%%%%%%%%%%%%%%%%%%%%%%%%%%%%%%%%%%%%%%%%%%%%%%%%%%%%%%%%%%%%%%%%

\subsection{Checking the gauge fixing}
\label{subsec:ratio}

For each gauge-fixed configuration we also measured
the ratios
\be
r_{i} \,\equiv \, \frac{e_{i} }{ e_{2}}
\ee
with $\,i=1\mbox{,}\,4\,$ and $6\,$. For the cases
$\,i=1\,$ and $\,i=4\,$ this quantity is essentially
independent of the configuration and of the lattice size;
therefore, for each algorithm,
after averaging over all the
configurations, we take a final average over all the pairs
$(N, \beta)\,$.
The results are given in Table
\ref{table:ratios}. From that it is clear that the quantities
$\,e_{1}\,$, $\,e_{2}\,$ and $\,e_{4}$ not only decay with the
same rate (as said above) but also have
the same order of magnitude.

The situation for the ratio $\,e_{6}\,$ is considerably more complicated.
In fact, its value depends strongly not only on the
algorithm and on the lattice size $\,N\,$ but also on the
underlying configuration\footnote{~That this should be the case
was, somehow, expected since $\,e_{6}\,$ is a
less ``local'' quantity than $\,e_{1}\,$, $\,e_{2}\,$
and $\,e_{4}\,$, and therefore it represents a more sensible check
for the lattice Landau gauge condition \reff{eq:diverg0}.
}
$\left\{ U_{\mu}(\bx) \right\}$.
Namely, this quantity fluctuates so much that
if the average is taken over all the gauge-fixed configurations,
at a fixed lattice size, the corresponding standard deviation
is often comparable in magnitude to the average itself.
As an example, see Table \ref{table:ratios2} where we show our results
for all the methods on a $32^{2}\,$ lattice.
From these data (see also Figures \ref{fig:decays} and
\ref{fig:decays2})
it is clear that the Fourier acceleration method achieves
a much faster decay for $\,e_{6}\,$
than the Los Alamos method, the Cornell method and
the overrelaxation method. Actually this was expected. In fact,
by using one of these three local methods it can be easily
checked that, even when the condition \reff{eq:stop}
is satisfied, the quantities $\,Q_{\nu}\,$ [defined in
\reff{eq:charges}] are usually not constant but show a kind
of long-wavelength spatial fluctuation\footnote{~See also Figure 12
in \cite{GGKPSW} and Figure 1 in \cite{BPS}.}. The Fourier acceleration
method treats all the wavelength in the same way and therefore
it is not surprising that it is very effective in reducing
these spatial fluctuations. More surprising is the good
performance of the stochastic relaxation method which, although a local
method, also appears to be very efficient in reducing the fluctuations
of $\,Q_{\nu}\,$. Why this happens is not clear to us.

\begin{table}
\addtolength{\tabcolsep}{-1.0mm}
\hspace*{-1.0cm}
\protect\footnotesize
\begin{center}
\begin{tabular}{|| l | c | c ||}
\hline
\hline
algorithm & $ r_{1} $ & $ r_{4} $ \\
\hline
\hline
Los Alamos & $ 0.2445 \pm 0.0008 $ &
               $ 0.5197 \pm 0.0113 $ \\ \hline
Cornell & $ 0.1490 \pm 0.0054 $ &
     $ 4.5212 \pm 0.5858 $ \\ \hline
overrelaxation & $ 0.1061 \pm 0.0106 $ & 
             $ 3.2357 \pm 0.3482 $ \\ \hline
stochastic overr. & $ 0.0151 \pm 0.0039 $ &
               $ 0.9642 \pm 0.0214 $ \\ \hline
Fourier acceleration & $ 0.0483 \pm 0.0148 $ &
                      $ 1.0219 \pm 0.0374 $ \\ \hline
\hline
\end{tabular}
\end{center}
\caption{~The ratios $\,r_{1}\,$ and $\,r_{4}\,$ for each
          algorithm. Averages are taken first over the gauge-fixed
          configurations and then over the different pairs
          $(N, \beta)\,$. The error bars are one standard deviation.}
\label{table:ratios}
\end{table}

\begin{table}
\addtolength{\tabcolsep}{-1.0mm}
\hspace*{-1.0cm}
\protect\footnotesize
\begin{center}
\begin{tabular}{|| l | c | c | c ||}
\hline
\hline
algorithm & $ r_{6} $ & $ \min r_{6} $ & $ \max r_{6} $ \\
\hline
\hline
Los Alamos & $ 10.7 \times 10^6 \pm 4.9 \times 10^6 $ & $ 4939.7 $ &
               $ 2.23 \times 10^9 $ \\ \hline
Cornell & $ 19.3 \times 10^6 \pm 19.0 \times 10^6 $ & $ 86.3 $ &
     $ 1.90 \times 10^9 $ \\ \hline
overrelaxation & $ 2.4 \times 10^5 \pm 1.2 \times 10^5 $ & $ 34.1 $ &
             $ 1.1 \times 10^7 $ \\ \hline
stochastic overr. & $ 3.9 \times 10^3 \pm 2.0 \times 10^3 $ & $ 4.93 $ &
               $ 1.8 \times 10^5 $ \\ \hline
Fourier acceleration & $ 5.6 \times 10^3 \pm 4.5 \times 10^3 $ & $ 0.84 $ &
                      $ 4.5 \times 10^5 $ \\ \hline
\hline
\end{tabular}
\end{center}
\caption{~The ratio $\,r_{6}\,$ (average, minimum and maximum
          value) for each
          algorithm. Averages are taken over the gauge-fixed
          configurations for $N = 32\,$ (for other lattice
          sizes the results are similar). The error bars
          are one standard deviation.}
\label{table:ratios2}
\end{table}

%%%%%%%%%%%%%%%%%%%%%%%%%%%%%%%%%%%%%%%%%%%%%%%%%%%%%%%%%%%%%%%%%%%%%%
%%%%%%%%%%%%%%%%%%%%%%%% TUNING OF THE ALGORITHMS %%%%%%%%%%%%%%%%%%%%
%%%%%%%%%%%%%%%%%%%%%%%%%%%%%%%%%%%%%%%%%%%%%%%%%%%%%%%%%%%%%%%%%%%%%%

\subsection{Discussion of the Tuning of the Algorithms}

Let us now discuss the tuning of the different methods. 
The values for the optimal choice of the various parameters,
for different pairs $(N, \beta)\,$, are reported in Tables
\ref{Table.Cor}--\ref{Table.FFT}. An estimate
of their uncertainties is also indicated. From our
simulations we noticed
that a good tuning becomes more and more important
as the lattice size increases. At small lattice sizes, in fact,
the relaxation time $\tau$ displays a kind of plateau
around the minimum while, as $N$ increases, the
absolute minimum becomes more and more pronounced. The uncertainties
indicated in these tables are therefore slightly under-estimated
for the smaller lattice sizes, and slightly over-estimated for
the larger lattice sizes.
In Figure \ref{fig:tuning},
as an example, we show typical graphs of our tuning parameters
for some of the larger values of
$N\,$, done at the ``third level'' of tuning (see Section 6).

\vskip 4mm

We now try to verify, using our data, the expressions
suggested in Section 5 for the various tuning parameters.

In order to check the relation (\ref{eq:omegafit}) for the
optimal choice of the
overrelaxation parameter $\omega\,$, we rewrote that equation as
\be
\frac{2\,-\,\omega_{opt}}{\omega_{opt}} \, = \, \frac{C_{opt}}{N}
\label{eq:omegafit2}
\ee
and fitted our results to find a value for the constant $C_{opt}\,$.
After discarding the
datum for $N = 8$, we obtained $C_{opt}\, =\, 1.53 \pm
0.35\,$.
In Figure \ref{fig:copt} we show, together, the data and
the fitting curve.

For the parameter $\alpha$ of the Cornell method we do not
have a simple formula as \reff{eq:omegafit2} but we have conjectured,
in Section \ref{section:Tuning}, a relation between
$\omega_{opt}$ and $\alpha_{opt}$. Namely we suggested
[see formulae (\ref{eq:omegainCorn}) and (\ref{eq:Nstima})]
\be
\omega_{opt} \, = \, \alpha_{opt} \, 2\,d\,\left(\,1\,-\,{\cal E}_{stat}
\,\right)
\; \mbox{,}
\ee
where $d$ is the dimension of the lattice and ${\cal E}_{stat}$ is
the value of the minimizing function at the minimum.
By using the data reported in Tables
\ref{Table.Cor} and \ref{Table.Overre} we plotted together, in
Figure \ref{fig:omegavsalfa}, both sides of this equation. The
agreement is clearly good.

Finally we checked the two relations
introduced in Section \ref{section:Tuning} between $\omega$
and the tuning parameter $p$ for the stochastic overrelaxation
method. In particular we plotted, in Figure \ref{fig:omegavsp},
the two ratios
\be
\frac{\omega_{opt}\,-\,1}{p_{opt}}
 \qquad \mbox{and} \qquad
\frac{\left(\,\omega_{opt}\,-\,1\,\right)^{2}}{p_{opt}}
\ee
as a function of $\,N\,$. If one of these relations
[ see formulae \reff{eq:omegainp} and \reff{eq:omegainp2} ] is
correct we should obtain, for the corresponding ratio, a
constant value $1$.
From our data it is not
possible to reach a definite conclusion, but the first hypothesis,
namely
\be
\frac{\omega_{opt}\,-\,1}{p_{opt}} \; = \; 1
\;\mbox{,}
\ee
seems to be slightly better verified.

\begin{figure}[p]
\vspace*{0cm} \hspace*{-0cm}
\begin{center}
\epsfxsize = 4.0in
\leavevmode\epsffile{tuning_ove.ps} \\
\vspace*{5mm}
\epsfxsize = 4.0in
\leavevmode\epsffile{tuning_sto.ps}
\end{center}
\vspace*{-5mm}
\caption{~Plot
of the third level of tuning respectively for
(a) the overrelaxation method, at lattice size 20, and
(b) the stochastic overrelaxation method, at lattice size 28.
Error bars are one standard deviation. Note that points are correlated,
since the same 100 configurations are used for each value of the tuning
parameter.
}
\label{fig:tuning}
\end{figure}

\begin{figure}
\epsfxsize=0.4\textwidth
\centerline{\epsffile{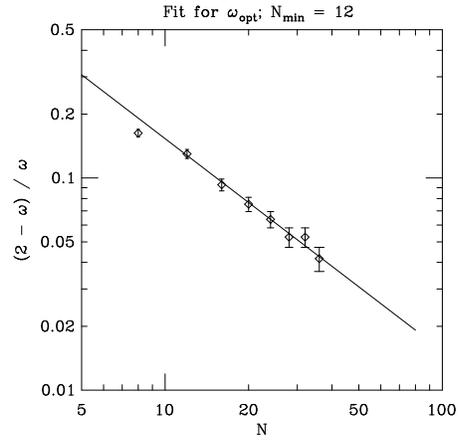}}
\caption{~Plot of the ratio $\,(2 - \omega_{opt}) / \omega_{opt}\,$
   (symbol: $\diamond$)
   as a function of $N\,$. The solid line is the fitting curve
   $\,C_{opt} / N\,$ with $\,C_{opt} = 1.53$.
}
\label{fig:copt}
\end{figure}

\begin{figure}
\epsfxsize=0.4\textwidth
\centerline{\epsffile{omegavsalfa.ps}}
\caption{~Plot of $\,\omega_{opt}\,$ (symbol: $\diamond$)
   and $\,\alpha_{opt}\, {\cal N}$ (symbol: $\Box$)
   as a function of $N$ with $\,{\cal N}\,$ given by equation
   \protect\reff{eq:Nstima}.
}
\label{fig:omegavsalfa}
\end{figure}

\begin{figure}
\epsfxsize=0.4\textwidth
\centerline{\epsffile{omegavsp.ps}}
\caption{~Plot of $\,(\omega_{opt} - 1) / p_{opt}\,$ (symbol: $\diamond$)
   and $\,(\omega_{opt} - 1)^2 / p_{opt}$ (symbol: $\Box$)
   as a function of $N$. In order to check the two hypotheses
   introduced in Section 5, the constant curve $1$ is
   also shown.
}
\label{fig:omegavsp}
\end{figure}

%%%%%%%%%%%%%%%%%%%%%%%%%%%%%%%%%%%%%%%%%%%%%%%%%%%%%%%%%%%%%%%%%%%%%%
%%%%%%%%%%%%%%%%%%%%%%% COMPUTATIONAL COST %%%%%%%%%%%%%%%%%%%%%%%%%%%
%%%%%%%%%%%%%%%%%%%%%%%%%%%%%%%%%%%%%%%%%%%%%%%%%%%%%%%%%%%%%%%%%%%%%%

\subsection{Computational Cost of the Algorithms}

To check the computational cost of the algorithms
we estimated the CPU time $T_{gf}$ necessary to update a single
site variable $g(\bx)$ by using the {\tt fortran} function {\tt mclock}.
As expected, the four local methods have very similar values for
$T_{gf}\,$ and essentially independent of the volume.
In particular we found
$T_{gf}\approx 9 \mu s$ for the
Los Alamos method, $T_{gf}\approx 9.5 \mu s$ for the
Cornell method, $T_{gf}\approx 11.5 \mu s$ for the overrelaxation
method and $T_{gf}\approx 10.5 \mu s$ for the stochastic
overrelaxation method.

\begin{figure}
\epsfxsize=0.4\textwidth
\centerline{\epsffile{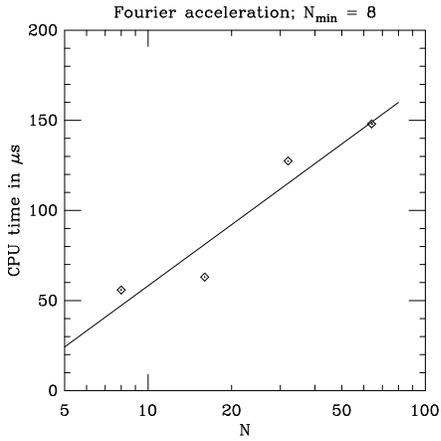}}
\caption{~Plot
   of CPU time $T_{gf}$ as a function of $N$ for the
   Fourier acceleration method. The solid line represents a
   least-squares fit to the ansatz $\,T_{gf} = A\,\log N \,+\, B\,$,
   with $\,A\,=\,48.91\,\pm\, 0.11\,$ and
   $\,B\,=\,-54.42\,\pm\,0.45\,$ (both in microseconds).
}
\label{fig:TGF}
\end{figure}

For the case of the Fourier acceleration method
$T_{gf}$ should increase \cite{PTVF} as $\log N\,$. We did a
least-squares fit of our data to the ansatz
\be
T_{gf}\, = \, A\,\log N \,+\, B
\label{eq:fitTGF}
\ee 
and we obtained $\,A\,=\,48.91\,\pm\, 0.11\,$ and
$\,B\,=\,-54.42\,\pm\,0.45\,$, both measured in 
microseconds.
In Figure \ref{fig:TGF} we show the points and the fitting curve.
For our range of lattice sizes, $\,T_{gf}\,$ for the Fourier acceleration
varied from $56\,\mu s$ to $148\,\mu s\,$. Of course, this 
``loss'' in efficiency, with respect to the local algorithms,
has to be taken into account when
the computational cost of this algorithm is analyzed.
In fact, even though the Fourier acceleration method
succeeds in eliminating critical slowing-down, and therefore
is more advantageous than the improved
local method in terms of the number of sweeps $N_{sw}\,$
required to achieve gauge fixing\footnote{~From 
Tables \ref{Table.Cor}--\ref{Table.FFT}
it is clear that the number of sweeps $N_{sw}$
increases roughly linearly with the lattice size for the
improved local methods, while it remains essentially constant for
the Fourier acceleration method.}, its performance
is effectively better only
at very large lattices.

\begin{figure}
\epsfxsize=0.4\textwidth
\centerline{\epsffile{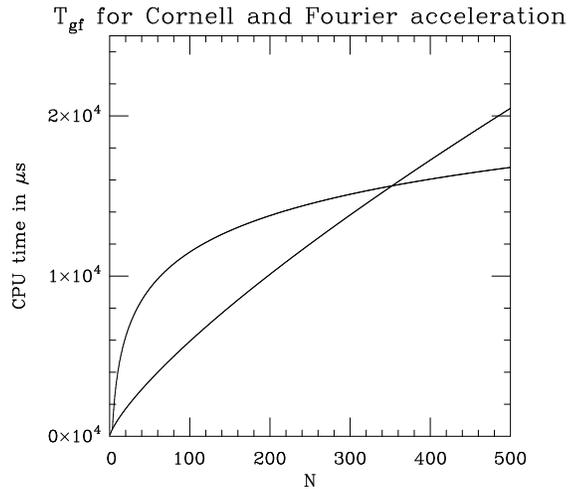}}
\caption{~Comparison
   between CPU times (divided by volume) as a function
   of the lattice size for the Cornell and the Fourier
   acceleration methods. The ``almost'' straight
   line corresponds to the Cornell method.
}
\label{plot:CPUtime}
\end{figure}

To illustrate this, let us compare the true CPU time
needed to gauge fix a configuration using Fourier acceleration and
our best improved local algorithm: the Cornell method. Since we
want to evaluate the total CPU time we have to look at
the number of sweeps $N_{sw}\,$, needed in average to achieve gauge fixing, as
a function of $N$. This quantity behaves in a manner similar to
the relaxation time $\tau\,$, namely it diverges with $N$
as a power of some dynamic critical exponent $\,z_{nsw}\,$.
This exponent should be similar to, but
strictly smaller than the exponent $z$ for the relaxation times.
In fact, $\,N_{sw}\,$ is a
quantity involving the behavior of the whole gauge-fixing process, and
therefore it includes faster modes (modes with ``smaller exponents'')
for the first few iterations, while
the relaxation time $\,\tau\,$ is
evaluated in the limit of large $t$, {\em i.e.}
when essentially only the slowest mode has survived.
The exponents $\,z_{nsw}\,$ for the various methods can be obtained,
together with the respective proportionality constants $\,c_{nsw}\,$,
from a fitting of the data in Tables \ref{Table.LosAlamos}--\ref{Table.FFT},
analogously to what was done for the $z$'s.
For the Cornell method we obtain $\,z_{nsw}\approx 0.77\,$
(which should be compared to $\,z\approx 0.82\,$ for the relaxation time)
and $c_{nsw}\approx 18\,$,
while for the Fourier acceleration we get $\,z_{nsw}\approx 0$ (as expected)
and $c_{nsw}\approx 67\,$.
From these estimates, the value $T_{gf}\approx 9.5 \mu s$ for the
Cornell method and the fit \reff{eq:fitTGF} for the Fourier
acceleration method,
we can get the
following approximate expressions for the time,
in microseconds, necessary to gauge fix a configuration:
\be
CPU time \;\approx\;
   \left\{ \begin{array}{ll}
               9.5 \, \left(\,18\,N^{0.77}\,\right) \,
     N^2 \, \mu s
                       & \quad \mbox{Cornell method} \\
      \phantom{ } & \phantom{ } \\
          \left(\,49 \log N \,-\, 54\,\right) \, 67 \,
      N^2 \, \mu s
          & \quad \mbox{Fourier acceleration}
        \end{array} \right.
\ee
In Figure \ref{plot:CPUtime} we show a plot of these two functions
(divided by the volume $\,N^2$).
Clearly,
Fourier acceleration becomes the method of choice only
at lattice sizes $N$ of order of $350\,$ !!
Of course this analysis is very machine- and code-dependent.
In particular we remark that, at the present stage,
our code for the Cornell method has
been considerably optimized, while the one for the Fourier acceleration
can still be improved, hopefully increasing its
running speed by a constant
factor, which we think could be as high as 2. Moreover, if
we used  a condition on the quantity $e_{6}\,$
to stop the gauge fixing, instead of the one given in equation
\reff{eq:stop}, then the computational cost of the Cornell method
would increase much more than that of the Fourier acceleration method,
as is clear from the discussion in Section \ref{subsec:ratio}.
In any case, it seems unlikely that
the Fourier acceleration method would become the method
of choice at lattice sizes smaller than around 100 sites.

%%%%%%%%%%%%%%%%%%%%%%%%%%%%%%%%%%%%%%%%%%%%%%%%%%%%%%%%%%%%%%%%%%%%%%
%%%%%%%%%%%%%%%%%%%%%%%%%%%%%% CONCLUSION %%%%%%%%%%%%%%%%%%%%%%%%%%%%
%%%%%%%%%%%%%%%%%%%%%%%%%%%%%%%%%%%%%%%%%%%%%%%%%%%%%%%%%%%%%%%%%%%%%%

\subsection{Conclusions}

From our numerical simulations it is clear that the Fourier acceleration method
is very effective in reducing critical slowing-down for the problem of $SU(2)$
Landau gauge fixing in two dimensions.
On the other hand, its computational cost is much larger than that of the
improved local methods, and therefore an accurate analysis
should be always done to decide which method
to use. The result of this analysis,
as stressed in the previous subsection, will depend
on the code, on the machine and on the condition used to stop the
gauge fixing. From our data it seems that, at least up to lattice
size of order $100$, the improved local methods should be always preferred.

From the point of view of computational cost, the Cornell method
is clearly the best among the improved local methods.
However,
if the condition used to stop the gauge fixing is not
\reff{eq:stop}, this conclusion could be different. In particular we saw
that the stochastic overrelaxation is very effective in 
relaxing the value of the quantity $e_6\,$, defined in
\reff{eq:e6}.

All the improved methods, including the Fourier acceleration, have
the disadvantage of requiring tuning. However, the relations
for the overrelaxation, the Cornell and the stochastic
overrelaxation methods that were introduced in Section 5 and
checked in Section 7.3 make the tuning of these
methods simpler. Moreover, as reported in Section 6, the values of
$\tau$ for the improved methods [for a fixed pair $( N, \beta)\,$]
are much more stable than for the Los Alamos method. Therefore,
in order to find the optimal choice 
of tuning parameter within a few per cent,
it suffices to perform a few numerical tests.

Finally, from the discussion in Sections 4 and 7.2, it is clear that
the quantities $e_{1}$--$e_{5}$ are essentially equivalent
as a check of the goodness of the gauge fixing.
Namely, when one of these quantities is
measured, the evaluation of any of the others does not provide any
new information. On the contrary, the quantity $e_6\,$ represents
a more sensible check of the gauge-fixing condition and,
in our opinion, should always be evaluated.

In a future paper \cite{future} we will try to extend this
analysis to the more interesting case of lattice gauge
theory in four dimensions.

%%%%%%%%%%%%%%%%%%%%%%%%%%%%%%%%%%%%%%%%%%%%%%%%%%%%%%%%%%%%%%%%%%%%%%
%%%%%%%%%%%%%%%%%%%%%% ACKNOWLEDGMENTS %%%%%%%%%%%%%%%%%%%%%%%%%%%%%%%
%%%%%%%%%%%%%%%%%%%%%%%%%%%%%%%%%%%%%%%%%%%%%%%%%%%%%%%%%%%%%%%%%%%%%%

\section*{Acknowledgments}

The authors would like to thank M.Passera, A.Pelissetto, M.Schaden,
A.Sokal and D.Zwanziger for helpful discussions and suggestions.

%%%%%%%%%%%%%%%%%%%%%%%%%%%%%%%%%%%%%%%%%%%%%%%%%%%%%%%%%%%%%%%%%%%%%%
%%%%%%%%%%%%%%%%%%%%%%%%%% BIBLIOGRAPHY %%%%%%%%%%%%%%%%%%%%%%%%%%%%%%
%%%%%%%%%%%%%%%%%%%%%%%%%%%%%%%%%%%%%%%%%%%%%%%%%%%%%%%%%%%%%%%%%%%%%%

%%%%%%%%%%%%%%%%%%%%%%%%%%%%%%%%%%%%%%%%%%%%%%%%%%%%%%%%%%%%%%%%%%%%%%
%%%%%%%%%%%%%%%%%%%%%% TABLES AND FIGURES %%%%%%%%%%%%%%%%%%%%%%%%%%%%
%%%%%%%%%%%%%%%%%%%%%%%%%%%%%%%%%%%%%%%%%%%%%%%%%%%%%%%%%%%%%%%%%%%%%%

%
\clearpage
\begin{table}
\addtolength{\tabcolsep}{-1.0mm}
\hspace*{-1.0cm}
\protect\footnotesize
\begin{center}
\begin{tabular}{|| c | c | c ||}
\hline
\hline
$ N $ & $ \tau $ & sweeps \\
\hline
\hline
$ 8 $ & $ 14.71 \pm 0.68 $ &$  330 \pm  16 $ \\ \hline 
$ 12 $ & $ 29.87 \pm 1.14 $ &$  600 \pm  18 $ \\ \hline 
$ 16 $ & $ 53.32 \pm 2.00 $ &$ 1054 \pm  34 $ \\ \hline 
$ 20 $ & $ 86.55 \pm 3.17 $ &$ 1650 \pm  91 $ \\ \hline 
$ 24 $ & $ 125.01 \pm 7.32 $ &$ 2159 \pm  79 $ \\ \hline 
$ 28 $ & $ 157.22 \pm 5.56 $ &$ 2797 \pm 103 $ \\ \hline 
$ 32 $ & $ 205.21 \pm 8.03 $ &$ 3372 \pm 105 $ \\ \hline 
$ 36 $ & $ 282.78 \pm 14.29 $ &$ 4389 \pm 155 $ \\ \hline 
\hline
\end{tabular}
\end{center}
\caption{~The relaxation time and the number of sweeps
          for the Los Alamos method.
          Error bars are one standard deviation.}
\label{Table.LosAlamos}
\end{table}
\begin{table}
\addtolength{\tabcolsep}{-1.0mm}
\hspace*{-1.0cm}
\protect\footnotesize
\begin{center}
\begin{tabular}{|| c | c | c | c | c ||}
\hline
\hline
$ N $ & $ \tau $ & $ \alpha $ & sweeps & min. func. \\
\hline
\hline
$ 8 $ & $ 3.22 \pm 0.30 $ & $ 0.489 $ & $  86 \pm 5 $ & $ 0.1171 \pm 0.0012 $ \\ \hline 
$ 12 $ & $ 4.54 \pm 0.33 $ & $ 0.484 $ & $ 117 \pm 6 $ & $ 0.0688 \pm 0.0005 $ \\ \hline 
$ 16 $ & $ 6.31 \pm 0.63 $ & $ 0.481 $ & $ 152 \pm 9 $ & $ 0.0430 \pm 0.0003 $ \\ \hline 
$ 20 $ & $ 6.92 \pm 0.40 $ & $ 0.480 $ & $ 181 \pm 8 $ & $ 0.0296 \pm 0.0002 $ \\ \hline 
$ 24 $ & $ 8.60 \pm 0.53 $ & $ 0.482 $ & $ 217 \pm 8 $ & $ 0.0218 \pm 0.0001 $ \\ \hline 
$ 28 $ & $ 9.52 \pm 0.61 $ & $ 0.483 $ & $ 232 \pm 9 $ & $ 0.0168 \pm 0.0001 $ \\ \hline 
$ 32 $ & $ 10.70 \pm 0.43 $ & $ 0.484 $ & $ 260 \pm 7 $ & $ 0.0134 \pm 0.0001 $ \\ \hline 
$ 36 $ & $ 11.65 \pm 0.44 $ & $ 0.484 $ & $ 278 \pm 7 $ & $ 0.0108 \pm 0.0001 $ \\ \hline 
\hline
\end{tabular}
\end{center}
\caption{~The relaxation time, the coefficient $ \alpha \,$,
	  the number of sweeps and
	  the value of the minimizing function
          for the Cornell method.
          Error bars are one standard deviation.
          The estimated error on the parameter $ \alpha $ is
          about $ 0.002 $ for all lattice sizes.}
\label{Table.Cor}
\end{table}
\begin{table}
\addtolength{\tabcolsep}{-1.0mm}
\hspace*{-1.0cm}
\protect\footnotesize
\begin{center}
\begin{tabular}{|| c | c | c | c ||}
\hline
\hline
$ N $ & $ \tau $ & $ \omega $ & sweeps \\
\hline
\hline
$ 8 $ & $ 2.77 \pm 0.35 $ & $ 1.72 $ & $  77 \pm 9 $ \\ \hline 
$ 12 $ & $ 3.56 \pm 0.15 $ & $ 1.77 $ & $  95 \pm 3 $ \\ \hline 
$ 16 $ & $ 4.84 \pm 0.20 $ & $ 1.83 $ & $ 136 \pm 5 $ \\ \hline 
$ 20 $ & $ 6.51 \pm 0.32 $ & $ 1.86 $ & $ 208 \pm 32 $ \\ \hline 
$ 24 $ & $ 7.84 \pm 0.38 $ & $ 1.88 $ & $ 208 \pm 9 $ \\ \hline 
$ 28 $ & $ 9.04 \pm 0.52 $ & $ 1.90 $ & $ 249 \pm 11 $ \\ \hline 
$ 32 $ & $ 10.34 \pm 0.62 $ & $ 1.90 $ & $ 257 \pm 9 $ \\ \hline 
$ 36 $ & $ 12.33 \pm 0.61 $ & $ 1.92 $ & $ 306 \pm 7 $ \\ \hline 
\hline
\end{tabular}
\end{center}
\caption{~The relaxation time, the coefficient $ \omega $
          and the number of sweeps for the overrelaxation method.
          Error bars are one standard deviation.
          The estimated error on the parameter $ \omega $ is
          about $ 0.01$ for all lattice sizes.}
\label{Table.Overre}
\end{table}
\begin{table}
\addtolength{\tabcolsep}{-1.0mm}
\hspace*{-1.0cm}
\protect\footnotesize
\begin{center}
\begin{tabular}{|| c | c | c | c ||}
\hline
\hline
$ N $ & $ \tau $ & $ p $ & sweeps \\
\hline
\hline
$ 8 $ & $ 3.35 \pm 0.28 $ & $ 0.64 $ & $  96 \pm  5 $ \\ \hline 
$ 12 $ & $ 4.38 \pm 0.18 $ & $ 0.72 $ & $ 129 \pm  4 $ \\ \hline 
$ 16 $ & $ 6.48 \pm 0.46 $ & $ 0.78 $ & $ 189 \pm 11 $ \\ \hline 
$ 20 $ & $ 7.84 \pm 0.30 $ & $ 0.84 $ & $ 234 \pm  7 $ \\ \hline 
$ 24 $ & $ 9.47 \pm 0.46 $ & $ 0.85 $ & $ 281 \pm 13 $ \\ \hline 
$ 28 $ & $ 11.28 \pm 0.67 $ & $ 0.88 $ & $ 331 \pm 13 $ \\ \hline 
$ 32 $ & $ 12.67 \pm 0.53 $ & $ 0.89 $ & $ 366 \pm 12 $ \\ \hline 
$ 36 $ & $ 14.95 \pm 0.87 $ & $ 0.90 $ & $ 421 \pm 15 $ \\ \hline 
\hline
\end{tabular}
\end{center}
\caption{~The relaxation time, the coefficient $ p $
          and the number of sweeps for the stochastic
          overrelaxation method.
          Error bars are one standard deviation.
          The estimated error on the parameter $ p $ is
          about $ 0.02 $ for all lattice sizes.}
\label{Table.Stoc}
\end{table}
\begin{table}
\addtolength{\tabcolsep}{-1.0mm}
\hspace*{-1.0cm}
\protect\footnotesize
\begin{center}
\begin{tabular}{|| c | c | c | c ||}
\hline
\hline
$ N $ & $ \tau $ & $ \alpha $ & sweeps \\
\hline
\hline
$ 8 $ & $ 3.08 \pm 0.33 $ & $ 0.193  $ & $  80 \pm 7 $ \\ \hline 
$ 16 $ & $ 3.30 \pm 0.34 $ & $ 0.170  $ & $  84 \pm 7 $ \\ \hline 
$ 32 $ & $ 3.11 \pm 0.21 $ & $ 0.160  $ & $  92 \pm 8 $ \\ \hline 
$ 64 $ & $ 3.46 \pm 0.33 $ & $ 0.160 $ & $  93 \pm 9 $ \\ \hline 
\hline
\end{tabular}
\end{center}
\caption{~The relaxation time, the coefficient $ \alpha \,$
         and the number of sweeps
         for the Fourier acceleration method.
         Error bars are one standard deviation.
         The estimated error on the parameter $ \alpha $ is
         about $ 0.003 $ for all lattice sizes.}
\label{Table.FFT}
\end{table}
\clearpage
\begin{table}
\addtolength{\tabcolsep}{-1.0mm}
\protect\footnotesize
\begin{center}
\begin{tabular}{||r|r@{\ $\pm$ \ }r| r@{\ $\pm$ \ }r|
  r@{\ (}r@{\mbox{\ DF, level\ $=$\ }}r||} \hline \hline
 $N_{min}$   & \multicolumn{2}{c|}{$z$} &
\multicolumn{2}{c|}{$c$} & \multicolumn{3}{c||}{$\chi^2$} \\ \hline
\hline
    8 & 1.950 & 0.032 & 0.2441 & 0.0235 & 6.177 & 
 6 & 40.365 \%) \\ 
\hline
   {\bf 12 } & {\bf 1.986 } & {\bf 0.042 } & {\bf 0.2174 } & {\bf 0.0284 } & {\bf 4.443 } & 
 {\bf 5 } & {\bf 48.751 \%)} \\ 
\hline
   16 & 1.965 & 0.060 & 0.2332 & 0.0448 & 4.196 & 
 4 & 38.017 \%) \\ 
\hline
   20 & 1.919 & 0.090 & 0.2727 & 0.0810 & 3.718 & 
 3 & 29.353 \%) \\ 
\hline
   24 & 2.030 & 0.171 & 0.1857 & 0.1082 & 3.131 & 
 2 & 20.900 \%) \\ 
\hline
   28 & 2.281 & 0.238 & 0.0779 & 0.0636 & 0.819 & 
 1 & 36.551 \%) \\ 
\hline
   32 & 2.722 & 0.543 & 0.0164 & 0.0313 & 0.000 & 
 0 & 100.000 \%) \\ 
\hline
\hline
\end{tabular}
\end{center}
\caption{~Weighted
least-squares fit for $\tau = c \,
 N^{z}$ at $N^2/\beta = 32$,
using lattice sizes $N \geq N_{min}$, for the Los Alamos
method. Errors 
represent one standard deviation, and ``DF'' stands for
``degrees of freedom''.
Confidence level is the probability that $\chi^2$ would
equal or exceed the observed value.
The line in boldface marks our preferred fit.
}
\label{zfit_losal}
\end{table}

\begin{table}
\addtolength{\tabcolsep}{-1.0mm}
\protect\footnotesize
\begin{center}
\begin{tabular}{||r|r@{\ $\pm$ \ }r| r@{\ $\pm$ \ }r|
  r@{\ (}r@{\mbox{\ DF, level\ $=$\ }}r||} \hline \hline
% \multicolumn{8}{||c||}{ Fit for the Cornell Method } \\ \hline
 $N_{min}$   & \multicolumn{2}{c|}{$z$} &
\multicolumn{2}{c|}{$c$} & \multicolumn{3}{c||}{$\chi^2$} \\ \hline
\hline
    8 & 0.854 & 0.049 & 0.5509 & 0.0877 & 1.134 & 
 6 & 98.002 \%) \\ 
\hline
   12 & 0.849 & 0.062 & 0.5611 & 0.1156 & 1.114 & 
 5 & 95.284 \%) \\ 
\hline
   {\bf 16} & {\bf 0.825} & {\bf 0.089} & {\bf 0.6088} & {\bf 0.1833} & {\bf 0.977} & 
{\bf  4} & {\bf 91.329 \%)} \\ 
\hline
   20 & 0.859 & 0.106 & 0.5409 & 0.1940 & 0.608 & 
 3 & 89.456 \%) \\ 
\hline
   24 & 0.762 & 0.168 & 0.7605 & 0.4401 & 0.045 & 
 2 & 97.759 \%) \\ 
\hline
   28 & 0.788 & 0.283 & 0.6936 & 0.6878 & 0.032 & 
 1 & 85.749 \%) \\ 
\hline
   32 & 0.721 & 0.471 & 0.8809 & 1.4627 & 0.000 & 
 0 & 100.000 \%) \\ 
\hline
\hline
\end{tabular}
\end{center}
\caption{~Weighted
least-squares fit for $\tau = c \,
 N^{z}$ at $N^2/\beta = 32$,
using lattice sizes $N \geq N_{min}$, for the Cornell
method. Errors 
represent one standard deviation, and ``DF'' stands for
``degrees of freedom''.
Confidence level is the probability that $\chi^2$ would
equal or exceed the observed value.
The line in boldface marks our preferred fit.
}
\label{zfit_cor}
\end{table}
\begin{table}
\addtolength{\tabcolsep}{-1.0mm}
\protect\footnotesize
\begin{center}
\begin{tabular}{||r|r@{\ $\pm$ \ }r| r@{\ $\pm$ \ }r|
  r@{\ (}r@{\mbox{\ DF, level\ $=$\ }}r||} \hline \hline
 $N_{min}$   & \multicolumn{2}{c|}{$z$} &
\multicolumn{2}{c|}{$c$} & \multicolumn{3}{c||}{$\chi^2$} \\ \hline
\hline
    8 & 1.094 & 0.045 & 0.2390 & 0.0329 & 3.381 & 
 6 & 75.974 \%) \\ 
\hline
   12 & 1.120 & 0.048 & 0.2205 & 0.0325 & 1.062 & 
 5 & 95.745 \%) \\ 
\hline
   {\bf 16 } & {\bf  1.120 } & {\bf  0.069 } & {\bf  0.2202 } & {\bf  0.0485 } & {\bf  1.061 } & {\bf  4 } & {\bf  90.034 \%)} \\ 
\hline
   20 & 1.063 & 0.108 & 0.2673 & 0.0949 & 0.577 & 
 3 & 90.157 \%) \\ 
\hline
   24 & 1.101 & 0.164 & 0.2340 & 0.1296 & 0.479 & 
 2 & 78.696 \%) \\ 
\hline
   28 & 1.239 & 0.301 & 0.1445 & 0.1513 & 0.185 & 
 1 & 66.707 \%) \\ 
\hline
   32 & 1.491 & 0.659 & 0.0590 & 0.1375 & 0.000 & 
 0 & 100.000 \%) \\ 
\hline
\hline
\end{tabular}
\end{center}
\caption{~Weighted
least-squares fit for $\tau = c \,
 N^{z}$ at $N^2/\beta = 32$,
using lattice sizes $N \geq N_{min}$, for the overrelaxation
method. Errors 
represent one standard deviation, and ``DF'' stands for
``degrees of freedom''.
Confidence level is the probability that $\chi^2$ would
equal or exceed the observed value.
The line in boldface marks our preferred fit.
}
\label{zfit_ove}
\end{table}
\begin{table}
\addtolength{\tabcolsep}{-1.0mm}
\protect\footnotesize
\begin{center}
\begin{tabular}{||r|r@{\ $\pm$ \ }r| r@{\ $\pm$ \ }r|
  r@{\ (}r@{\mbox{\ DF, level\ $=$\ }}r||} \hline \hline
 $N_{min}$   & \multicolumn{2}{c|}{$z$} &
\multicolumn{2}{c|}{$c$} & \multicolumn{3}{c||}{$\chi^2$} \\ \hline
\hline
    8 & 1.048 & 0.043 & 0.3395 & 0.0449 & 3.673 & 
 6 & 72.089 \%) \\ 
\hline
{\bf 12 } & {\bf 1.086 } & {\bf 0.050 } & {\bf 0.3002 } & {\bf 0.0465 } & {\bf 1.330 } & {\bf 5 } & {\bf 93.178 \%)} \\ 
\hline
   16 & 1.034 & 0.081 & 0.3566 & 0.0940 & 0.680 & 
 4 & 95.377 \%) \\ 
\hline
   20 & 1.065 & 0.098 & 0.3217 & 0.1029 & 0.355 & 
 3 & 94.929 \%) \\ 
\hline
   24 & 1.084 & 0.171 & 0.3017 & 0.1751 & 0.338 & 
 2 & 84.455 \%) \\ 
\hline
   28 & 1.115 & 0.329 & 0.2702 & 0.3084 & 0.325 & 
 1 & 56.842 \%) \\ 
\hline
   32 & 1.407 & 0.609 & 0.0966 & 0.2062 & 0.000 & 
 0 & 100.000 \%) \\ 
\hline
\hline
\end{tabular}
\end{center}
\caption{~Weighted
least-squares fit for $\tau = c \,
 N^{z}$ at $N^2/\beta = 32$,
using lattice sizes $N \geq N_{min}$, for the stochastic overrelaxation
method. Errors 
represent one standard deviation, and ``DF'' stands for
``degrees of freedom''.
Confidence level is the probability that $\chi^2$ would
equal or exceed the observed value.
The line in boldface marks our preferred fit.
}
\label{zfit_sto}
\end{table}
\begin{table}
\addtolength{\tabcolsep}{-1.0mm}
\protect\footnotesize
\begin{center}
\begin{tabular}{||r|r@{\ $\pm$ \ }r| r@{\ $\pm$ \ }r|
  r@{\ (}r@{\mbox{\ DF, level\ $=$\ }}r||} \hline \hline
 $N_{min}$   & \multicolumn{2}{c|}{$z$} &
\multicolumn{2}{c|}{$c$} & \multicolumn{3}{c||}{$\chi^2$} \\ \hline
\hline
 {\bf 8 } & {\bf 0.036 } & {\bf 0.064 } & {\bf 2.8513 } & {\bf 0.6062 } & {\bf 0.751 } & {\bf 2 } & {\bf 68.703 \%)} \\ 
\hline
   16 & 0.040 & 0.102 & 2.8080 & 1.0094 & 0.748 & 
 1 & 38.712 \%) \\ 
\hline
   32 & 0.157 & 0.169 & 1.8020 & 1.1286 & 0.000 & 
 0 & 100.000 \%) \\ 
\hline
\hline
\end{tabular}
\end{center}
\caption{~Weighted
least-squares fit for $\tau = c \,
 N^{z}$ at $N^2/\beta = 32$,
using lattice sizes $N \geq N_{min}$, for the Fourier acceleration
method. Errors 
represent one standard deviation, and ``DF'' stands for
``degrees of freedom''.
Confidence level is the probability that $\chi^2$ would
equal or exceed the observed value.
The line in boldface marks our preferred fit.
}
\label{zfit_fft}
\end{table}
\clearpage

\begin{figure}[p]
\begin{center}
\vspace*{0cm} \hspace*{-0cm}
\epsfxsize=0.4\textwidth
\leavevmode\epsffile{dataz.loc.ps}
\hspace{2cm}
\epsfxsize=0.4\textwidth
\epsffile{dataz.cor.ps}  \\
\vspace*{0.5cm}
\epsfxsize=0.4\textwidth
\leavevmode\epsffile{dataz.ove.ps}
\hspace{2cm}
\epsfxsize=0.4\textwidth
\epsffile{dataz.sto.ps}  \\
\vspace*{0.5cm}
\epsfxsize=0.4\textwidth
\centerline{\epsffile{dataz.fft.ps}}
\end{center}
\vspace{-0.5cm}
\caption{~Power-law fit for
     (a) the Los Alamos method,
     (b) the Cornell method,
     (c) the overrelaxation method,
     (d) the stochastic overrelaxation method and
     (e) the Fourier acceleration method.
}
\label{fig:dataz}
\end{figure}

\end{document}